\documentclass[a4paper,11pt]{article}
\pdfoutput=1
\usepackage{jheppub}
\usepackage{graphicx}
\usepackage{subcaption}
\usepackage{dsfont}
\usepackage[dvipsnames]{xcolor}
\usepackage{etoolbox}
\usepackage[T1]{fontenc}
\usepackage{amsmath}
\usepackage{mathtools}
\usepackage{amsthm}
\usepackage{physics}
\makeatletter
\patchcmd{\maketitle}{\@fpheader}{\\}{}{}

\makeatother

\usepackage{babel}

\title{Multi-mouth Traversable Wormholes}
\author[a,b]{Roberto Emparan,}
\author[c,d]{Brianna Grado-White,}
\author[c]{Donald Marolf,}
\author[b,e]{and Marija Toma\v{s}evi\'c}

\affiliation[a]{Instituci\'o Catalana de Recerca i Estudis Avan\c cats (ICREA),
Passeig Llu\'{\i}s Companys 23, E-08010 Barcelona, Spain}
\affiliation[b]{Departament de F{\'\i}sica Qu\`antica i Astrof\'{\i}sica, Institut de Ci\`encies del Cosmos,
Universitat de Barcelona, Mart\'{\i} i Franqu\`es 1, E-08028 Barcelona, Spain}
\affiliation[c]{Department of Physics, University of California, Santa Barbara, CA 93106, USA}
\affiliation[d]{Martin Fisher School of Physics, Brandeis University, Waltham MA, USA}
\affiliation[e]{Kavli Institute for Theoretical Physics
University of California, Santa Barbara, CA 93106}

\emailAdd{emparan@ub.edu}
\emailAdd{bgradowhite@brandeis.edu}
\emailAdd{marolf@physics.ucsb.edu}
\emailAdd{mtomasevic@icc.ub.edu}

\abstract{We describe the construction of traversable wormholes with multiple mouths in four spacetime dimensions and discuss associated quantum entanglement. Our wormholes may be traversed between any pair of mouths. In particular, in the three-mouth case they have fundamental group $F_2$ (the free group on two generators). By contrast, connecting three regions $A,B,C$ in pairs ($AB$, $BC$, and $AC$) using three separate wormholes would give fundamental group $F_3$.  Our solutions are asymptotically flat up to the presence of possible magnetic fluxes or cosmic strings that extend to infinity. The construction begins with a two-mouth traversable wormhole supported by backreaction from quantum fields. Inserting a sufficiently small black hole into its throat preserves traversability between the original two mouths. This black hole is taken to be the mouth of another wormhole connecting the original throat to a new distant region of spacetime. Making the new wormhole traversable in a manner similar to the original two-mouth wormhole provides the desired causal connections. From a dual field theory point of view, when AdS asymptotics are added to our construction, multiparty entanglement may play an important role in the traversability of the resulting wormhole.}

\begin{document}

\maketitle

\section{Introduction}

Under certain natural assumptions, topological censorship theorems forbid constructions of traversable wormholes in classical general relativity \cite{Friedman:1993ty,Galloway:1999bp}.  In globally hyperbolic spacetimes obeying the null curvature condition, such theorems require causal curves to be deformable to curves that lie entirely in the boundary of the spacetime, and that these curves can be made to remain causal throughout the deformation. Recently, however, it was shown that well-controlled quantum effects could be used to violate the null energy condition in a manner allowing the construction of
traversable wormholes \cite{Gao:2016bin,Maldacena:2018lmt,Maldacena:2018gjk, Fu:2018oaq,Caceres:2018ehr,Marolf:2019ojx,Fu:2019vco} that circumvent these theorems; see also \cite{Horowitz:2019hgb, Maldacena:2019ufo} for studies of the dynamical production of such traversable wormholes.

\begin{figure}[!htbp]
        \begin{center}
        \begin{subfigure}{0.49\linewidth}
        \centering
                \includegraphics[width=0.85\textwidth]{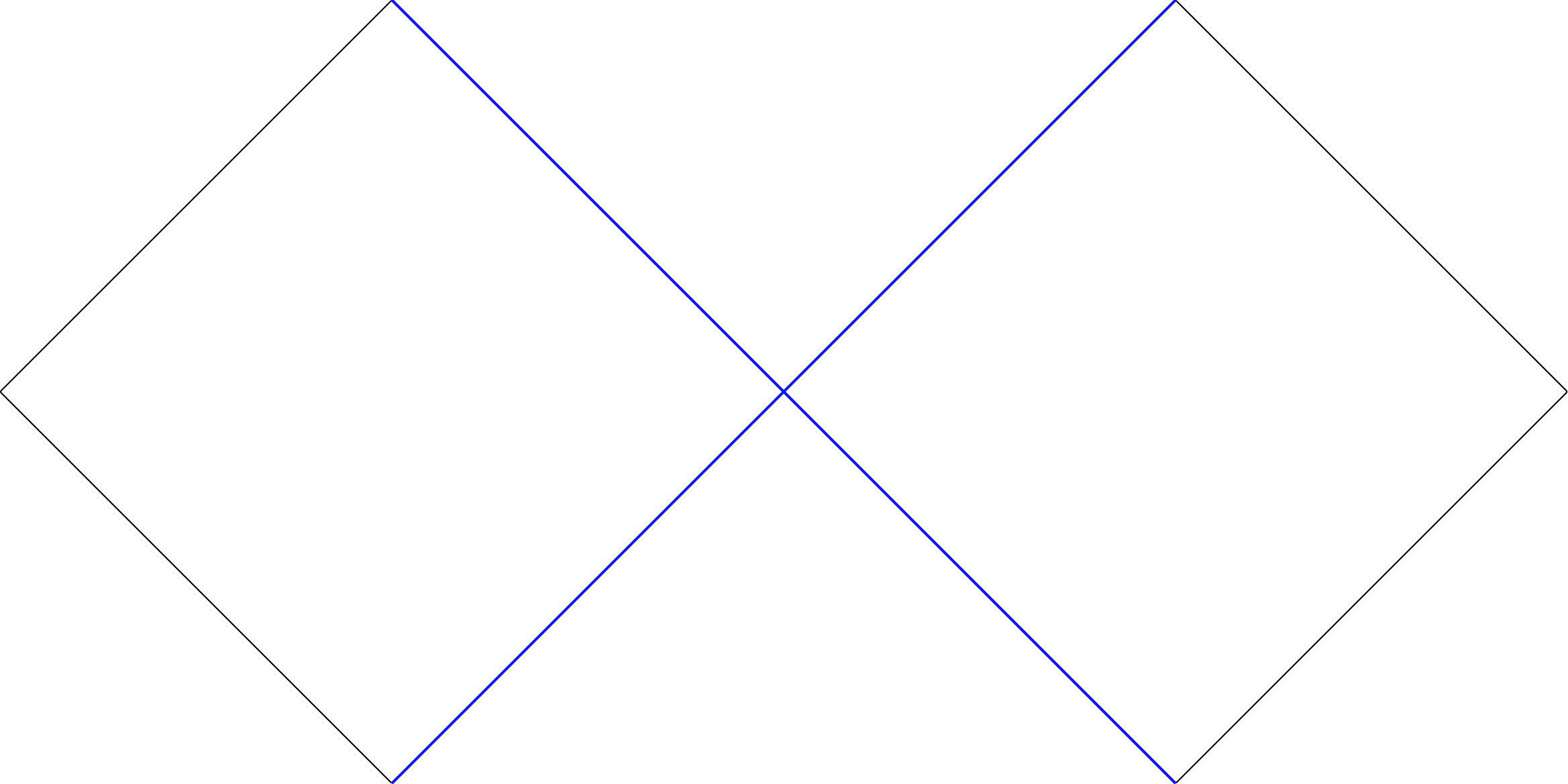}\quad
                \end{subfigure}
		\begin{subfigure}{0.49\linewidth}
        \centering
                \includegraphics[width=0.95\textwidth]{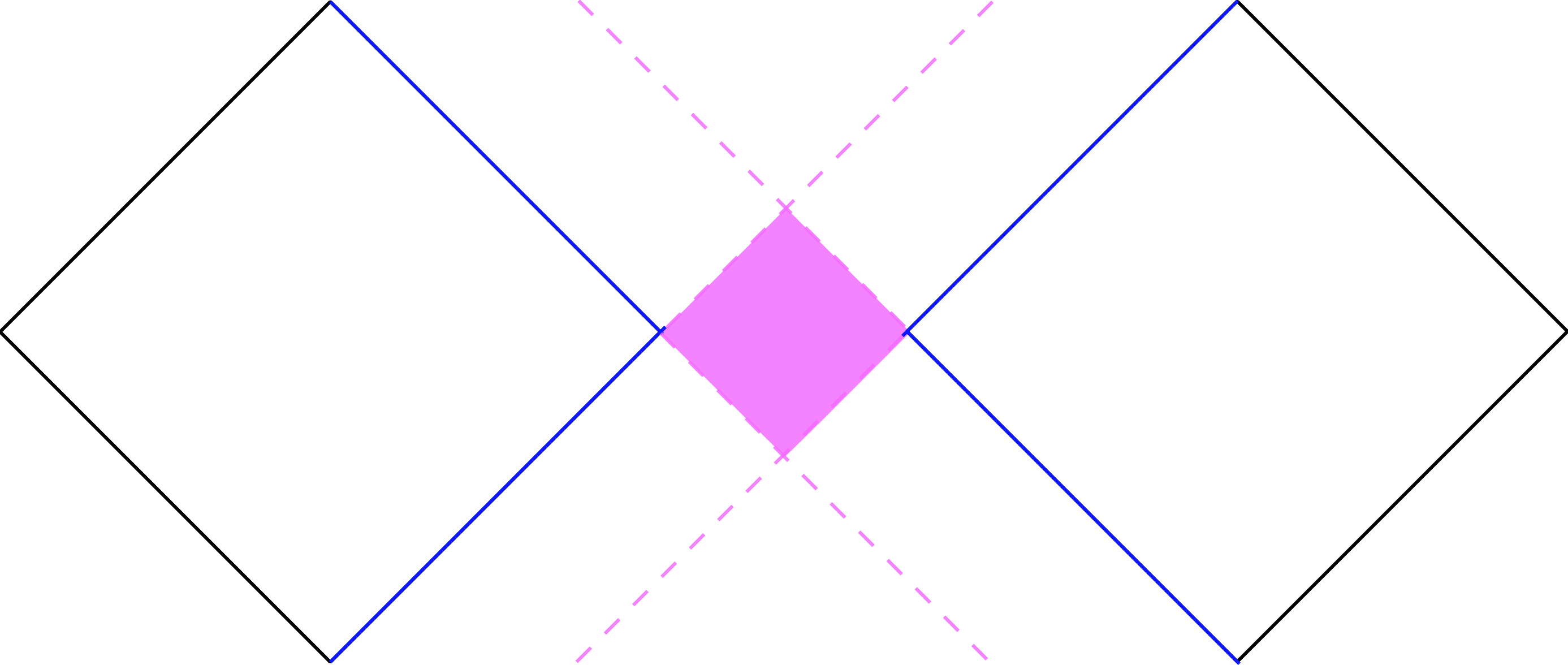}
                \end{subfigure}
        \end{center}
          \caption{\small {\bf Left:} A bifurcate horizon in a two-sided asymptotically flat spacetime.
          {\bf Right:} A spacetime with a causal shadow (shaded in purple). }
        \label{fig:bif}
\end{figure}

Quantum effects in gravity are typically difficult to control unless they are in some sense small. For this reason, the above constructions of traversable wormholes can be thought of as starting with background spacetimes that contain an almost traversable wormhole that can be rendered traversable with small corrections.  In classical solutions satisfying the null energy condition, this generally requires the background to contain a bifurcate horizon having no causal shadow\footnote{A causal shadow is defined as a bulk region which is causally disconnected from the boundary, see \cite{Headrick_2014} for more details.}\footnote{Though the wormholes of \cite{Maldacena:2018gjk} are not explicitly written in this form, \cite{Fu:2019vco} gave a similar construction which could be written as a perturbation around a bifurcate horizon.}; see Fig.~\ref{fig:bif}. Naively then, it might seem as if traversable wormholes are constrained to connect only two regions of spacetime having a single mouth in each region, as backgrounds with more interesting connectivity require some sort of finite causal shadow, and this in turn necessitates a larger amount of negative energy to make the wormhole traversable.

Nevertheless, we show below that constructions with higher connectivity can still be controlled.~Our analysis begins with the more familiar two-mouth asymptotically flat wormholes of \cite{Maldacena:2018gjk}, enhanced by including a large number $N_f$ of four-dimensional massless fermions. We then perturb this solution by adding a small black hole to the bottom of the wormhole throat. Wormholes are very fragile, and semiclassical black holes have large masses in Planck units, so one may worry that the insertion of this small black hole could destroy traversability.  However, the extreme redshift deep in the wormhole throat allows semiclassical black holes to sit in the bottom and leave  traversability intact. Indeed, we show that one can actively pass a small black hole through a wormhole mouth and place it at the bottom of the throat without choking it.

\begin{figure}[!htbp]
        \begin{center}
        \begin{subfigure}{0.54\linewidth}
        \centering
                \includegraphics[width=0.85\textwidth]{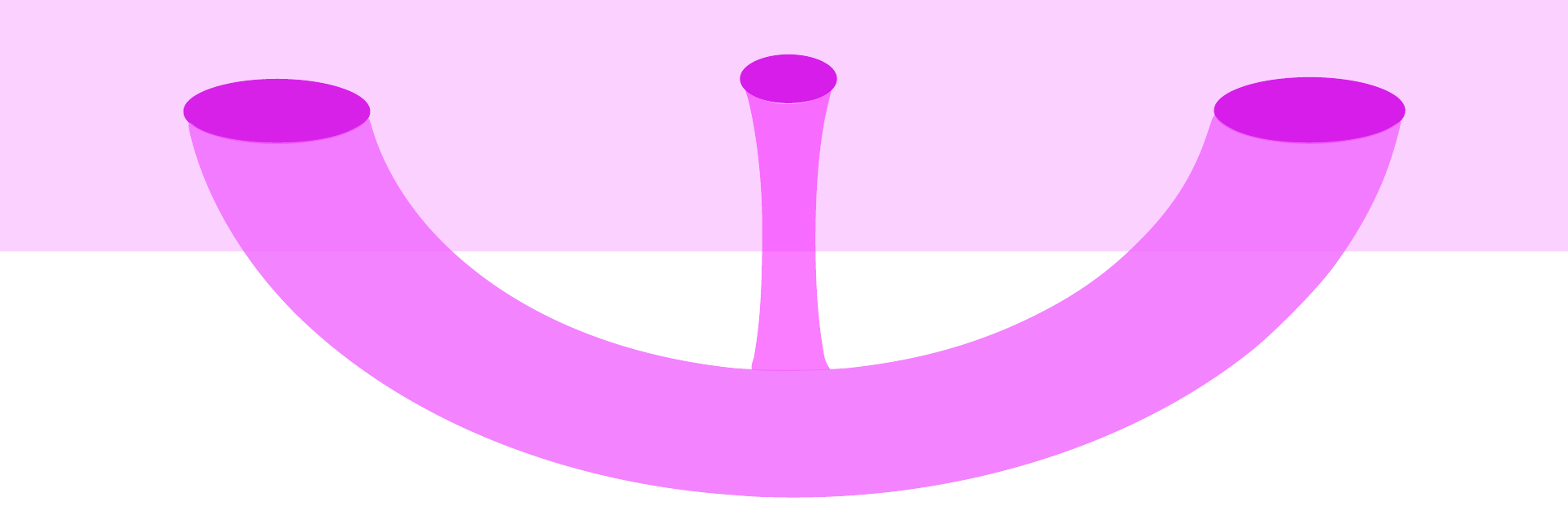}\quad
                \end{subfigure}
		\begin{subfigure}{0.44\linewidth}
        \centering
                \includegraphics[width=0.85\textwidth]{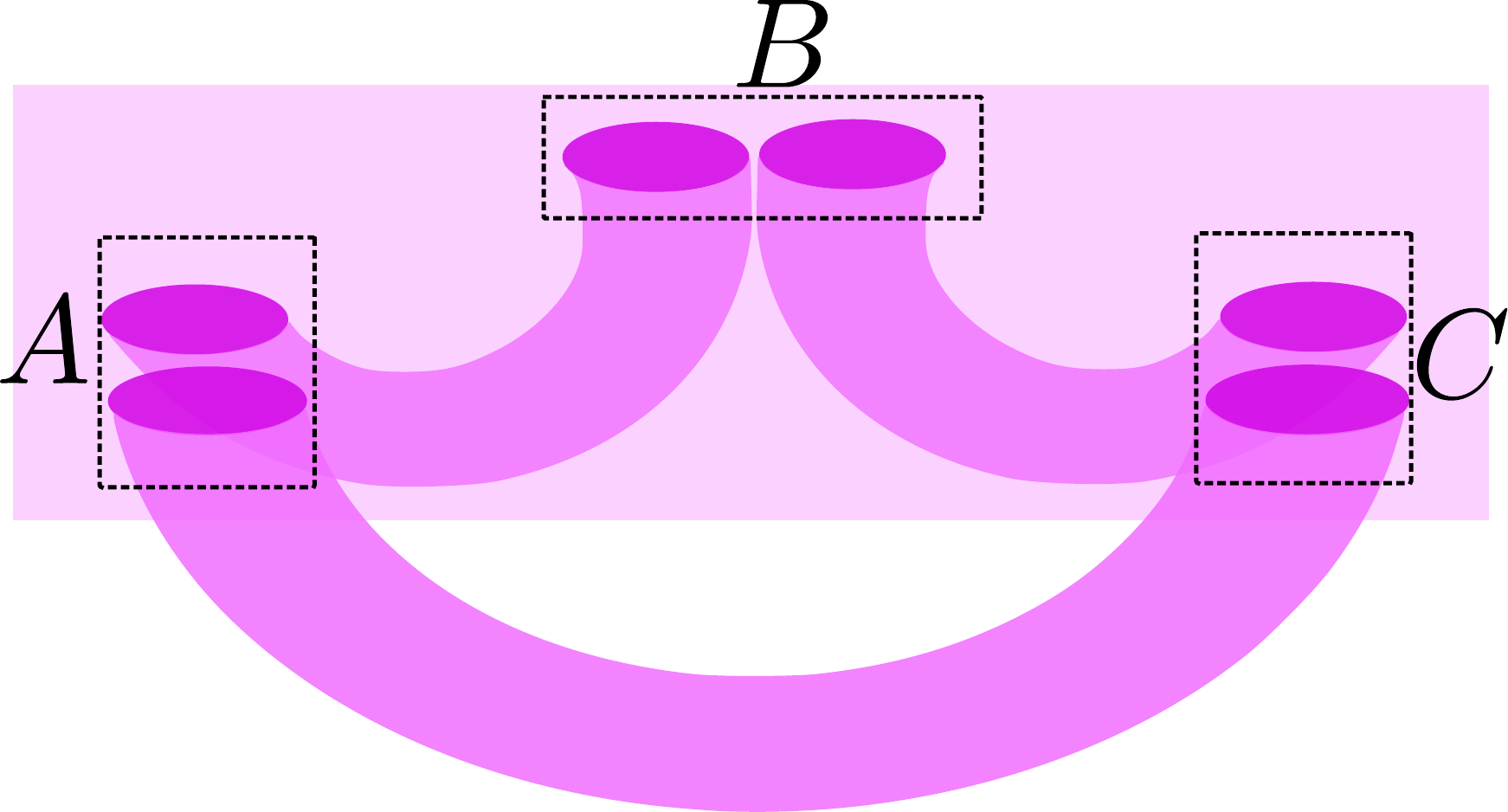}
                \end{subfigure}
        \end{center}
        \caption{\small {\bf Left:} A  two-dimensional analogue of our spatial topology has two handles. The actual  three-dimensional space has fundamental group $F_2$, the free group on two generators. {\bf Right:} A space with three wormholes connecting regions $A,B,C$ in pairs $AB$, $BC$, $AC$ has three handles. In three dimensions, the fundamental group would be $F_3$.}
        \label{fig:F23}
\end{figure}

We take this small black hole to contain an additional wormhole that connects to another distant region of spacetime. This new wormhole can then be made traversable with further quantum effects in a manner similar to the original, two-mouth wormhole. The resulting spacetime then has fundamental group $F_2$, the free group on two generators. This differs from the fundamental group $F_3$ that would be obtained by adding three separate two-mouth wormholes connecting three distant regions of spacetime $A,B,C$  in pairs $AB$, $BC$, and $AC$; see Fig.~\ref{fig:F23}.

The above construction also has interesting implications for the quantum states of wormholes.  First, the ability to add a small black hole to a two-mouth traversable wormhole indicates additional traversable excited states beyond those  anticipated in the analyses of \cite{Maldacena:2018lmt,Maldacena:2018gjk}. Second, at least when embedded in AdS/CFT, our three-mouth traversable wormhole appears to involve a new entanglement structure different from the TFD-like entanglement associated with two-mouth wormholes.

The paper is organized as follows. In Sec.~\ref{sec:review}, we review the construction  of four-dimensional asymptotically flat traversable wormholes of \cite{Maldacena:2018gjk}. We describe the gravitational construction of our multi-mouth traversable wormholes in Sec.~\ref{sec:grav}. We conclude with a brief discussion in Sec.~\ref{sec:disc}, focusing on quantum states and entanglement. The details of the matched asymptotics expansion for the construction that employs the wormholes of \cite{Maldacena:2018gjk}  are given in the Appendix \ref{appendix:mae}. In Appendix \ref{appendix:pertwhs}, we discuss the construction of multi-mouth traversable wormholes starting from the asymptotically flat two-mouth wormholes of \cite{Fu:2019vco}.

\section{Review of two-mouth traversable wormholes}
\label{sec:review}

We review here the construction of two-mouth traversable wormholes in four-dimensional asymptotically flat space of \cite{Maldacena:2018gjk}. As described in Appendix \ref{appendix:pertwhs},  the implementation of our construction of multi-mouth wormholes using the two-mouth asymptotically flat traversable wormholes of \cite{Fu:2019vco} proves to be more difficult.

Building a traversable wormhole requires some source of average negative null energy, so that $\int d \lambda T_{ab} k^a k^b < 0 $ for $k$ null and $\lambda$ an affine parameter. Intuitively, this is because null rays moving into the wormhole throat initially converge, but need to diverge to exit the other side. This focusing/defocusing of light rays is controlled by the null energy through the Raychaudhuri equation, with positive null energy causing null rays to focus and negative null energy causing them to defocus. Classical matter, which obeys the null energy condition $ T_{ab} k^a k^b > 0 $, is thus insufficient to construct a traversable wormhole. Quantum effects, however, can give rise to negative null energy  (e.g. Casimir energy) when certain boundary conditions are imposed on quantum fields, as can happen in the presence of non-trivial topology.

The wormholes of \cite{Maldacena:2018gjk} start with near-extremal magnetically charged Reissner-Nordstr\"{o}m (RN) black holes and take the near-horizon limit, where the metric becomes
\begin{equation}
\label{eq:NHL}
        ds^2 = r_e^2\left(-(\rho_r^2 - 1)d\tau_r^2 + \frac{d\rho_r^2}{\rho_r^2 - 1} + d\Omega^2\right).
\end{equation}
Here $r_e$ is the horizon radius of extremal RN black holes of charge $Q$ (quantized and dimensionless), which is 
\begin{equation}\label{reQ}
    r_e=\sqrt{\pi} l_P\frac{Q}{g}\,,
\end{equation}
where $g$ is the coupling constant of the $U(1)$ gauge field and 
\begin{equation}
    l_P=\sqrt{G_N}
\end{equation}
is the Planck length.
The coordinates $\rho_r$ and $\tau_r$ are functions of the usual $r$ and $t$ coordinates of RN. This near-horizon metric is AdS$_2 \times S^2$ with the AdS$_2$ factor presented in Rindler coordinates.  

Note, however, that the $S^2$ factor has constant size in \eqref{eq:NHL}. To make a traversable wormhole that connects different portions of asymptotically flat space,  the metric must be modified to allow the size of the $S^2$ to vary, so that the black hole spacetime can be sewn onto the asymptotically flat region. One may take this variation to be slow, with the perturbation away from AdS$_2 \times S^2$ being of the form
\begin{equation}
\label{eq:addphi}
    ds^2 = r_e^2\left(-(1 + \rho^2 + \gamma)d\tau^2 + \frac{d\rho^2}{1 + \rho^2 + \gamma} + (1 + \phi)d\Omega^2\right),
\end{equation}
where $\phi$ and $\gamma$ are small functions that, respectively, encode the changing size of the sphere and a perturbation of the AdS$_2$ factor. Now $\tau$ and $r$ are global coordinates for the AdS$_2$ factor, which make it appear easy to send causal signals from one side to the other.

From the Raychaudhuri equation, spacetimes of the form \eqref{eq:addphi} satisfying the null energy condition must have $\phi(\rho)$ monotonic.  But connecting the throat region to the asymptotic regions at both ends requires that the spheres $S^2$ grow at both the wormhole mouths, and therefore $\phi$ must grow in both directions at large $|\rho|$. Completing the construction in a solution of Einstein-Hilbert gravity thus requires the introduction of negative energy.

\begin{figure}[h]
        \begin{center}
        \begin{subfigure}{\linewidth}
        \centering
              \includegraphics[width=0.4\textwidth]{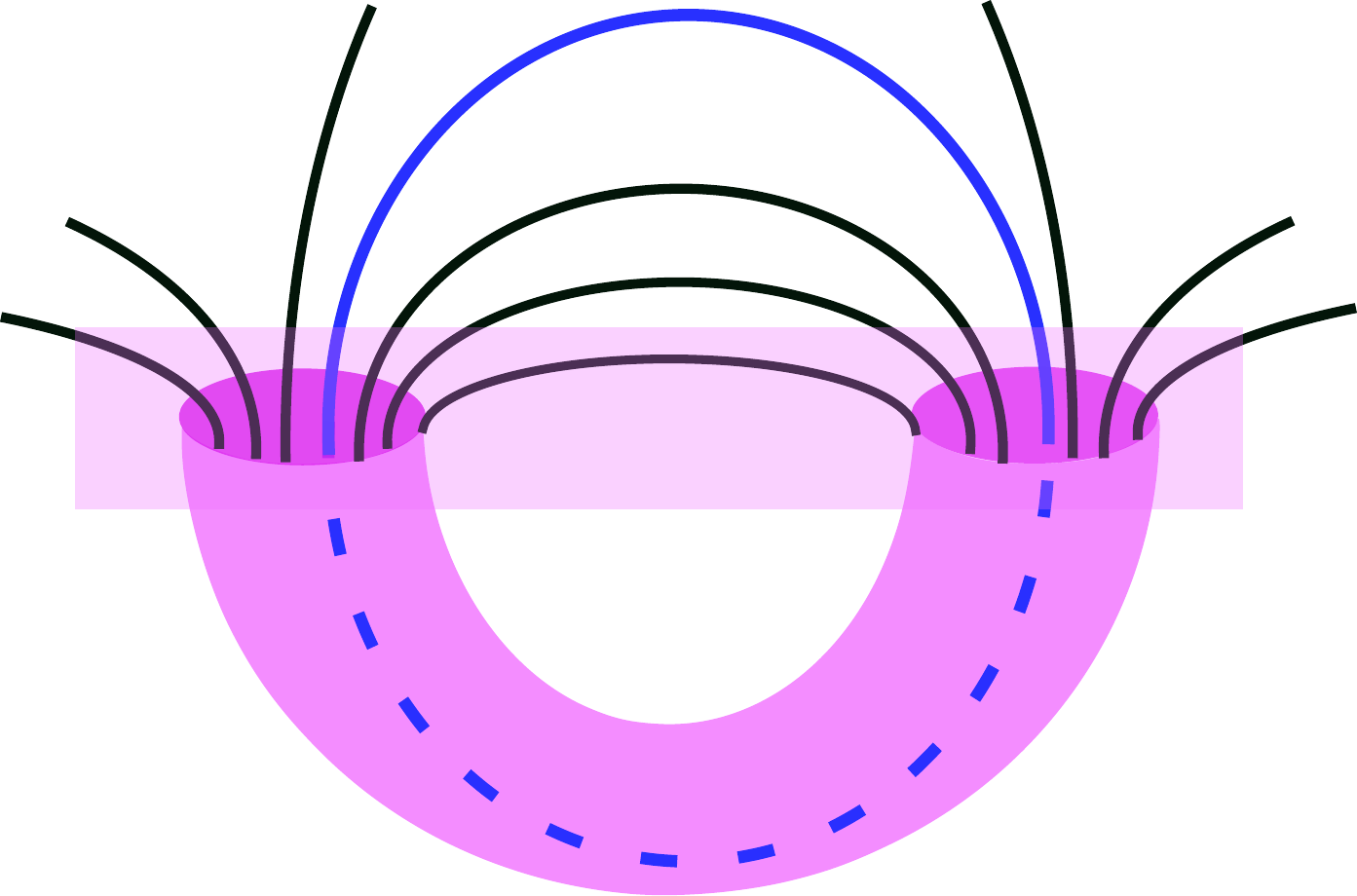}\quad
                \end{subfigure}
        \end{center}
        \caption{\small The traversable wormhole of \cite{Maldacena:2018gjk}. Magnetic field lines thread the wormhole, while fermions localize into their lowest Landau level near each field line.  For field lines that form closed loops (for example, in blue), this creates  effective 1+1 dimensional massless theories on $S^1 \times {\mathbb R}$ whose Casimir energy makes the wormhole traversable. }
        \label{fig:4d_mag}
\end{figure}

The construction of \cite{Maldacena:2018gjk} creates negative Casimir energy by using the magnetic field of the black hole and a massless, charged fermion field.  The magnetic field creates localized Landau levels near each field line, which gives a large number $Q$ of effective 1+1 dimensional massless fermions. As shown in Fig.~\ref{fig:4d_mag}, field lines that loop through the wormhole yield 1+1 dimensional theories on $S^1 \times {\mathbb R}$. Since constant $\phi$ yields the exact solution \eqref{eq:NHL} with vanishing stress-energy, a small negative stress-energy suffices to allow growth of $\phi$ at large positive and negative $\rho$ so long as the negative stress-energy threads the entire wormhole and this growth is correspondingly slow.

Ref.~\cite{Maldacena:2018gjk} showed that the energy of such wormholes differs from the energy of two disconnected extremal black holes by the amount\footnote{This assumes a simplified model for fermion propagation in the exterior region; see \cite{Maldacena:2018gjk} for details.} 
\begin{equation}\label{whenergy}
E=\frac{r_e^3}{G_N \ell^2}-\frac{N_f Q}{6}\left(\frac{\pi}{\pi\ell+d_{out}}-\frac1{4\ell}\right)\,,
\end{equation}
where $\pi \ell$ is equal to the ``length'' of the throat in an appropriate conformally rescaled metric, $d_{out}$ is the separation between mouths in the exterior region, and $Q$ is the integer magnetic charge of the black holes.  

The first term in \eqref{whenergy} is the energy above extremality of a near-extremal RN black hole, and although it is a classical contribution, it can be small when $\ell$ is large. It can thus be offset by the second, quantum contribution from the Casimir energy.  Minimizing \eqref{whenergy} determines the equilibrium wormhole length. When $d_{out}\ll \ell$ one gets
\begin{equation}\label{whell}
    \ell=\frac{16 r_e^3}{G_N N_f Q}\sim \frac{Q^2 l_P}{N_f}
\end{equation}
so that the wormhole binding energy is
\begin{equation}\label{Emin}
    E_{min}=-\frac{N_f Q}{16 \ell}=-\frac{N_f^2 g^3}{256 \pi^{3/2} Q l_P}\,.
\end{equation}
This solution requires that $d_{out}\ll Q^2 l_P/N_f$, but it is also possible to find configurations with $d_{out}\gg\ Q^2 l_P/N_f$, in which the balance is achieved between the two quantum terms in \eqref{whenergy} with the classical energy being negligible. In this case energy minimization gives $\ell=d_{out}/\pi$. The binding energy is still $E_{min}\sim -N_f Q/\ell$ but this is now much smaller than in \eqref{Emin}, since $\ell$ is much larger than in \eqref{whell}.

We emphasize that the magnetic field lines must form closed loops in order to generate Casimir energy.  This requires that both wormhole mouths be placed in the same asymptotic region of spacetime.  As a result, the mouths attract each other gravitationally. As long as the initial separation $d_{out}$ between the mouths is sufficiently large, the wormhole will remain open for long enough to cross it before collapsing -- the time to merger is $\sim d_{out}^{3/2}$ (which is just the time for two point masses to collide starting from rest), while the the transit time along the throat is parametrically $O(d_{out})$.  However, additional structure can be added to create longer-lived traversable wormhole solutions. This can be achieved by introducing an external magnetic field (in GR, this would be a Melvin flux tube \cite{Melvin:1963qx}) tuned to keep the mouths apart, or by attaching cosmic strings that pull them (exact solutions exist for both mechanisms \cite{Emparan:1999au,Emparan:2001bb}). Alternatively, instead of balancing them into exact (though unstable) equilibrium, one can set the mouths into a long-lived Keplerian orbit around each other, as proposed in \cite{Maldacena:2018gjk}. Though this orbiting will cause the wormhole mouths to radiate gravitational and electromagnetic waves and hence eventually coalesce, the time scale for this to happen is $d_{out}^3$, and so again much longer than the time needed to traverse the wormhole when the wormhole mouths are sufficiently far apart. Additionally, as noted in \cite{Fu:2019vco}, it is possible to create a more complicated stable solution  by anchoring cosmic strings to some stable spherical shell at a finite distance.  This approach uses several strings to attach each black hole to the shell, with each string anchored to a different location on the shell.  Stability arises from the fact that the angles between the strings depend on the location of the black holes.

It is expected that the wormhole throat must be longer than the distance between the wormhole mouths, though the wormholes of \cite{Fu:2019vco} approximately saturate this bound in certain limits.
In $d>4$ this is a sharp bound that follows from, for example, the Generalized Second Law \cite{C:2013uza}, or in AdS/CFT, from boundary causality \cite{Gao:2000ga}. These statements prohibit wormholes from being the fastest causal curves between distant points, and indeed the travel time through wormholes of \cite{Maldacena:2018gjk} is longer by a factor of order unity.\footnote{However, in $d=4$ asymptotically flat spacetimes, the Shapiro time-delay associated with the wormhole mouths means that the fastest causal curve between two distant points always lies far from the center of mass. Thus, the sharp bounds mentioned above are always trivially satisfied, and a sharp, local bound is lacking for wormhole transit times. However,  it may be possible to derive sharper local bounds by considering either the quantum focusing conjecture \cite{Bousso:2015mna}, or by considering short wormhole's tendency to form time machines \cite{Visser:1995cc}.}

\section{Gravitational construction}
\label{sec:grav}

Given the two-mouth wormhole described above, our idea for constructing multi-mouth traversable wormholes is simple: place a small, near-extremal black hole in the throat of a larger-mouth wormhole, and extend it into a wormhole with another small mouth in the same asymptotic region as the larger mouths. Technically, the insertion of the two small mouths in the initial large wormhole solution is a straightforward (if possibly tedious) problem of matched asymptotic expansions: the small mouths can be treated as perturbations of, respectively, the throat and the asymptotic region, while the effects of the latter on the mouths are incorporated as tidal perturbations of the near-extremal Reissner-Nordstr\"om black hole. In Appendix \ref{appendix:mae}, we explain how to obtain the solution to this perturbation problem. For our purposes here, we simply need the lowest order in the matched asymptotic expansion, in which the backreaction of the mouths is neglected. We will only go beyond this order in Sec.~\ref{Malda}, where we calculate how the insertion of the small mouth modifies the energetics in the backreacted solution. 

Matching the geometry of the small mouths onto the background spacetime is thus a generic and unproblematic part of the construction, but there are other aspects that must be dealt with more carefully. One still fairly simple question is that of mechanical equilibrium (and possibly stability) of the new configuration. Actually, this arises at the first order in the matched asymptotic expansions, as we explain in Appendix~\ref{appendix:mae}. Another problem is how to achieve the negative energies that make the throats traversable. The answers to these questions vary depending on the details of the model we choose -- in other words, on the tools of which we avail ourselves for the construction. We may restrict ourselves to working within the same theory as \cite{Maldacena:2018gjk}, with only fields and matter available in the Standard Model (specifically, a Maxwell field and light fermions electrically coupled to it, in addition to gravity) -- or the Beyond-the-Standard-Model dark sector of \cite{Maldacena:2020sxe} -- to construct our multiboundary traversable wormholes. Or, instead, we may resort to a larger set of tools, as did \cite{Fu:2019vco} (using e.g. cosmic strings as may appear in, say, grand unified theories), and aim at a `proof of principle' that such wormholes are possible with reasonable matter and field content, e.g., satisfying basic energy conditions, and possibly within the landscape of string theory. Allowing only Standard Model tools of course makes the task more difficult.

\subsection{Multi-mouth wormhole construction guide}

We begin by asking how equilibrium can be achieved when one introduces a new wormhole mouth into the throat of the wormholes constructed in \cite{Maldacena:2018gjk}.
If near-extremal magnetic RN solutions approximately describe all three mouths in a magnetic field background, then equilibrium should not be hard to achieve.
A uniform magnetic field can be approximated by the field in between two large, static magnetic sources (even nonlinearly in GR \cite{Emparan:2001gm}). The third mouth can thus be thought of as sitting in a uniform, static magnetic field. This will push the
third mouth to one side, but the deep gravitational well can be used to make the forces on this mouth balance at a finite displacement.  In Appendix~\ref{appendix:mae}, we work out how this problem is solved when constructing the backreacted solution. Configurations with a small black hole in the throat in equilibrium can thus be found.

However, from a purely mechanical perspective, perhaps the simplest possibility is to let the small black hole be charged under a different $U(1)$ gauge field than the bigger mouths. This of course introduces physics beyond the Standard Model. Equilibrium configurations can then be found that preserve the natural reflection symmetry of the two-mouth solution, with the new source sitting in equilibrium at the bottom of the throat. 

We must also consider the external mouth to which the mouth in the throat connects.  This third external mouth will face stability issues similar to those described for the two-mouth wormhole above, which can then be resolved in similar ways. In particular, even if the configuration is unstable, it
can still be sufficiently long-lived to allow the throats to be traversed so long as the additional black hole remains small. This is so even though the addition of the small mouth in the throat will increase the time required to traverse the larger wormhole due to a Shapiro-like time delay that we will analyze in Sec.~\ref{between}.

Having described the mechanics involved with adding the third mouth, let us now move on to the problem of achieving negative Casimir energies that thread the associated wormhole. The effective two-dimensional massless fermions used to build the original two-mouth wormhole will still travel along magnetic field lines, which form loops along the non-contractible cycles of the wormhole and thus provide negative Casimir energies.  Some of these non-contractible field lines will thread the third mouth and hold it open as desired.

We get more varied possibilities if we enlarge our toolbox beyond the Standard Model. For instance, still using the magnetic line mechanism of \cite{Maldacena:2018gjk}, we can allow for three $U(1)$ gauge fields, and three flavors of fermions electrically coupled to each of the gauge fields. Then, with each pair of the mouths having opposite magnetic charges under one of the $U(1)$'s,\footnote{More precisely, the two big mouths can have charges $(Q_1,Q_2,0)$, $(-Q_1,0,Q_3)$ and the small one $(0,-Q_2,-Q_3)$, with $|Q_1|\gg |Q_2|, |Q_3|$. This also allows easily for symmetric equilibrium positions for the small mouth.} the fermions travel along field lines in an independent manner.

\begin{figure}[!htbp]
        \begin{center}
        \begin{subfigure}{0.6\linewidth}
        \centering
         \includegraphics[width=0.9\textwidth]{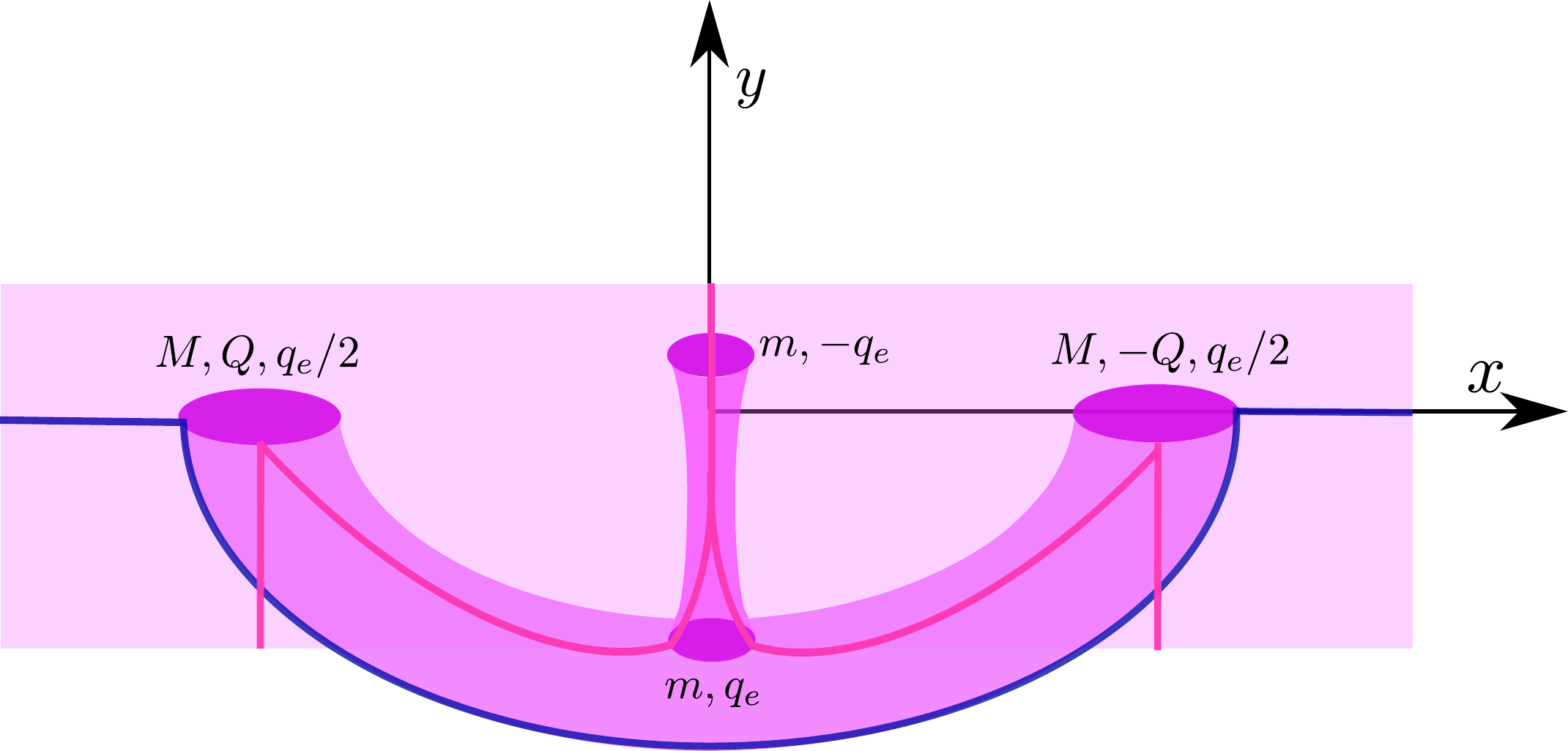}\quad
                \end{subfigure}
        \begin{subfigure}{0.39\linewidth}
        \centering
         \includegraphics[width=0.9\textwidth]{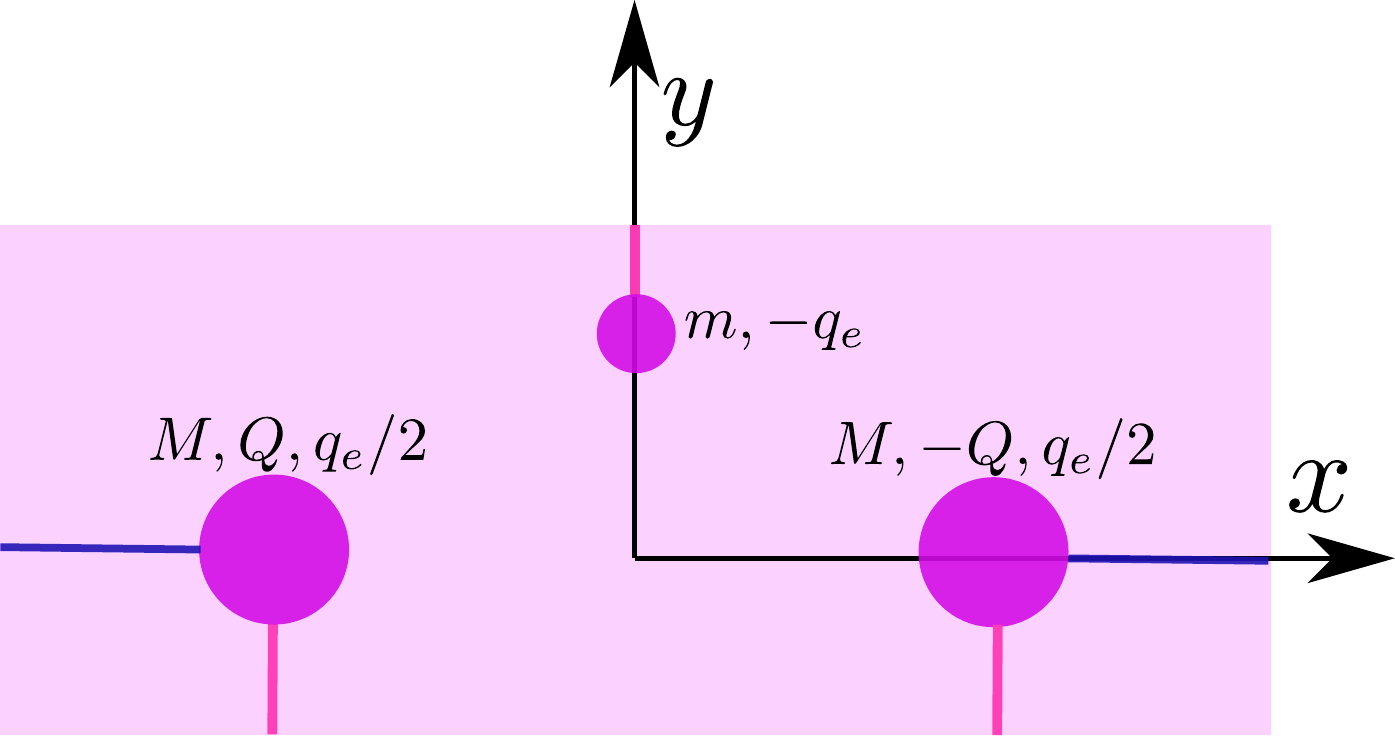}\quad
                \end{subfigure}
        \end{center}
        \caption{\small {\bf{Left}}: A three-mouth wormhole, held in mechanical equilibrium by cosmic strings. The cosmic strings needed for Casimir energy are omitted here for illustrational clarity, but would wrap the compact cycles around each pair of mouths. We take the small mouth to be charged under a different $U(1)$ symmetry than the previous black holes, with charge $q_e$. This is done to maintain the symmetry of the solution. Field lines from this small mouth in the throat will flow through the wormhole and exit through the large mouths, giving them each a charge $q_e/2$.  The other small mouth is  placed equidistant between the larger mouths, but off the axis connecting them, to avoid overlapping with the compact strings that are needed to generate Casimir energy (otherwise, the compact string would run down this small wormhole mouth, and not directly around the non-contractible cycle of the wormhole handle). We can then add two more cosmic strings that run in from infinity along the $y-$axis. The tension of these strings is equal, and can be chosen to compensate for the magnetic and gravitational forces. The tension of the original string that stretches along the $x-$axis can then be increased to compensate for the additional force on the large wormhole mouths from the small wormhole mouth. {\bf{Right}}: The top view.}
        \label{fig:3wh}
\end{figure}

Cosmic strings with zero modes traveling along loops of string provide the requisite Casimir energy in a simpler manner. We may use it in a hybrid fashion, by adding the third mouth to the magnetic-line model of \cite{Maldacena:2018gjk} and thread it with two cosmic strings, each separately linked to the two big mouths; or else, if that hybrid is deemed too ugly to regard, replace the fermions in the original two mouths with a cosmic string as in \cite{Fu:2019vco} in addition to the two new cosmic strings, one along each new cycle. An explicit example is given in Fig.~\ref{fig:3wh}.

Finally, notice that the assumption about a hierarchy of mouth sizes, i.e.,  very different charge numbers in the big and small mouths, can be achieved within the Standard Model, while keeping the mouth geometries semiclassical, given the large separation between the electroweak and Planck scales. However, we will see later that the energetics of the throat demand that the number of flavors $N_f$ be large in order to allow the insertion of the small mouth. Following \cite{Maldacena:2018gjk}, the Standard Model may provide as many as $N_f=54$.

Thus, while many potential constructions are possible, the cosmic string method of \cite{Fu:2019vco}, possibly augmented with additional gauge fields, serves to prove that it is possible to construct multi-mouth wormholes sufficiently long-lived to be traversable. Their existence within the Standard Model, following the methods of \cite{Maldacena:2018gjk}, also seems likely, even if its detailed investigation is more complicated.

\subsection{Size limits on the third mouth}
\label{Malda}

As described in \cite{Maldacena:2018gjk}, the traversability property is extremely fragile, as it can be destroyed by perturbations of a small-but-finite size. We should therefore study more carefully just how large the third mouth can be.  In particular, we should note that the semiclassical analysis used above requires the third mouth to be larger than the Planck length, indeed, larger than $\sqrt{N_f}l_P$, since a very large number of species $N_f$ restricts the validity of the semiclassical description to length scales $> \sqrt{N_f}l_P$. We should understand the conditions under which these constraints can be satisfied.

As reviewed in Sec.~\ref{sec:review}, from the standpoint of the Raychaudhuri equation, the key point is that the positive mass of the small mouth creates a focusing effect within the wormhole throat that counteracts the defocusing effect of the negative Casimir energy.  If the focusing effect of the small mouth is too large, the topological censorship arguments of \cite{Friedman:1993ty} require the throat to collapse and for traversability to be destroyed.
We expect that this places an upper bound on the mass of the small black hole such that its energy, as measured from outside the wormhole, does not exceed the binding energy of the wormhole. We now perform an analysis that confirms this expectation precisely. 

We start with the solution for the wormhole interior as described in \cite{Maldacena:2018gjk}, which we then perturb with a localized source for the small black hole deep inside the throat. As we have seen, the geometry in this region is described by a metric of the form of \eqref{eq:addphi}, where the small functions $\gamma$ and $\phi$ will be treated to linear order.
The Einstein equations are sourced by the Maxwell field stress-energy tensor, the Casimir energy from the fermions, and the small wormhole mouth. The former two contributions were computed in \cite{Maldacena:2018gjk}, and for the latter we introduce a number of useful simplifications.\footnote{See Appendix~\ref{appendix:mae} for how this relates to the construction of matched asymptotics.} First, since the additional wormhole mouth is small we can understand its backreaction by treating it as a localized, delta-function mass source. This is a codimension-three source, and although it is possible to perturbatively solve for its backreaction, we will simplify the problem further. We are interested in the effect of the source on the overall size of the $S^2$ along the throat -- which is controlled by the scalar field $\phi$ in \eqref{eq:addphi} -- and we can smear the source over the $S^2$, so it acts as a codimension-one, domain wall defect on the throat. This is a good approximation for studying the gravitational effect of the small mouth at distances in the throat larger than $r_e$. With this simplification, the geometry varies only along the throat direction $\rho$, and then $\phi$ and $\gamma$ are obtained by solving ordinary differential equations.

It will be sufficient to solve the $(\tau\tau)$ equation. To linear order in $\phi$ and $\gamma$ the Einstein tensor takes the form
\begin{equation}
     G_{\tau\tau} = \gamma - (1 + \rho^2)(-1 + \rho\phi' + (1 + \rho^2)\phi'') - (1 + \rho^2)\phi \label{tau}\,.
\end{equation}
The corresponding stress tensor consists of magnetic energy density, fermion Casimir energy, and localized mass source,
\begin{equation}
    T_{\tau\tau} = T^{\rm mag}_{\tau\tau} + T^{\rm fer}_{\tau\tau} + T^{\delta}_{\tau\tau},
\end{equation}
which are given by
\begin{equation}
\begin{aligned}
    T^{\rm mag}_{\tau\tau} =& -\frac{1}{4g^2}g_{\tau\tau}F^2 = \frac{1}{8\pi G_N}((1 + \rho^2)(1 - 2\phi) + \gamma),  \\ T^{\rm fer}_{\tau\tau} =& -\frac{\alpha}{8\pi G_N}\,, \qquad T^{\delta}_{\tau\tau} = \frac{\beta}{4\pi G_N} \delta(\rho)\,.
\end{aligned}
\end{equation}
We have placed the source at the center of the throat $\rho=0$. In the next section we will consider off-center positions (the off-center displacement created by the electric interaction between the charged black holes is discussed in Appendix~\ref{appendix:mae}). 
Furthermore we have defined\footnote{In order to get the dimensions of the source term correctly, bear in mind that $\rho$ is dimensionless and physical lengths are in units of $r_e$.}
\begin{equation}\label{alphabeta}
 \alpha = \frac{G_N Q N_f}{4\pi r_e^2}\,,\qquad   \beta = \frac{G_N m}{r_e}\,,
\end{equation}
with $m$ the mass of the small black hole. We have smeared it over the sphere $S^2$ of radius $r_e$, so we can trust the solution for $|\rho|$ larger than one, but not near $\rho=0$.

The Einstein $(\tau\tau)$ equation now involves only $\phi$ and not $\gamma$, and takes the form
\begin{equation}
    (1+\rho^2)\phi''+\rho\phi'-\phi=\frac{\alpha}{1+\rho^2}-2\beta\,\delta(\rho)\,.
\end{equation}
We solve this as
\begin{equation}\label{phisol}
    \phi(\rho) = \alpha(1 + \rho \arctan\rho) - \beta|\rho|\,.
\end{equation}
%
%
The contribution $\beta|\rho|$ from the small black hole is the gravitational potential that a mass creates in $1+1$ gravity, which is how gravity along the throat behaves over scales larger than its thickness $r_e$.  
As anticipated, we see the focusing effect of the mass $m$, which makes $\phi$ decrease as $|\rho|$ grows. If $\beta$ is large enough, then at distances $|\rho| >1$ (where our approximations hold) this effect could overcome the defocusing of the negative Casimir energy and the throat would close, as $\phi$ must increase towards the wormhole exits in order to connect to the asymptotic regions. Therefore we will limit $\beta$ to values such that $\phi$ grows for large  $|\rho|$. This gives
\begin{equation}\label{mbound}
    \beta< \frac{\pi}2\alpha\,,
\end{equation}
%
which tells us that the maximum mass of the small black hole that we can put in the wormhole is
\begin{equation}\label{mboundapp}
    m < \frac{N_f Q}{8r_e} \,.
\end{equation}

This $m$ is locally measured in the vicinity of the small black hole, deep within the throat, while the energy as measured outside the wormhole is redshifted by a factor of $r_e/\ell$, giving
\begin{equation}
    E_{bh} < \frac{N_f Q }{16 \ell}\,,
\end{equation}
that is,
\begin{equation}\label{nocollapse}
   E_{bh} < |E_{min}|
\end{equation}
where $E_{min}$ is the energy gap between traversable and non-traversable wormholes that we saw earlier in \eqref{Emin}. 

Writing the constraint \eqref{mboundapp} in the form
\begin{equation}\label{mbound2}
    m< \frac1{8\sqrt{\pi}}g N_f m_P\,,
\end{equation}
where the Planck mass is $m_P=1/l_P$, we see that at weak coupling $g$ we need $N_f\gg 1$ for the small mouth to remain semiclassical, with $m\gg m_P$. Indeed the actual semiclassicality condition when many fermion species are present, namely, $m\gg \sqrt{N_f} m_P$, can also be satisfied for  $N_f\gg 1$.

The bound on the size of the small mouth may be even more stringent than \eqref{nocollapse}, since we expect that the radius of the small mouth cannot exceed the throat radius, that is, we must have
\begin{equation}\label{rbound}
G_N m< r_e\,.
\end{equation}
This bound will be more stringent than \eqref{mboundapp} whenever
\begin{equation}
    r_e<\sqrt{\frac{N_f Q}{4\pi}} l_P\,,
\end{equation}
or equivalently when
\begin{equation}\label{sizebound}
    Q<\frac{g^2 N_f}{4\pi^2}\,.
\end{equation}
We need $Q/g\gg 1$, and actually $Q/g\gg \sqrt{N_f}$, in order for the black holes to be semiclassically valid (see \eqref{reQ}), and the coupling will naturally be $g\lesssim 1$. Still, \eqref{sizebound} allows situations where, due to the presence of a large number of fermions $N_f$, the small black hole reaches its maximal size while the original throat is still far from collapsing. The reason is that $N_f$ enhances the binding energy of the wormhole while not affecting the classical size relations \eqref{rbound}. Conversely, when \eqref{sizebound} does not hold, the binding energy is sufficiently small, such that a black hole much smaller than the throat radius can nevertheless be heavy enough to overwhelm the negative Casimir energy, and thus collapse the wormhole. In this case, the approximations that lead to \eqref{nocollapse} hold well and the bound can be regarded as accurate. 

From this analysis we conclude that, if traversability is to be preserved, the addition of the third mouth should not lift the energy of the system above that of two disconnected large extremal black holes. More generally, we expect that multi-mouth wormholes always have lower energy than a collection of disconnected extremal black holes. 

One may also ask about applying our construction to the perturbative two-mouth wormholes of \cite{Fu:2019vco}.  In that case, the wormhole only remains open for a short time, and so can only be traversed by causal curves that start early enough at past null infinity. The addition of the small mouth makes this restriction even more stringent. We analyze it in Appendix~\ref{appendix:pertwhs}, concluding that the bound on the mass of the small mouth is stronger than \eqref{nocollapse} by an additional factor of $\ell_p/r_e$.  Correspondingly, larger numbers of fermions $N_f$ are thus required for the third mouth to enter the semiclassical regime while preserving traversability of the original throat.

\subsection{Lowering the small mouth down the throat}
\label{subsec:lowering}

The previous subsection dealt with configurations with a small mouth that is already at the bottom of the wormhole throat. Now we want to investigate if it is indeed possible to lower a mass $m$ to that place, starting from a position near the big wormhole exit. Since the wormhole is fragile, we carefully lower the mass in a Geroch-like adiabatic process. Standing at the wormhole mouth, we attach the mass to a string which we slowly release into the wormhole, so that the system is in equilibrium at every moment while the mass is lowered. In contrast to the conventional Geroch process, when the small mass reaches the bottom of the wormhole it will be at its equilibrium position, and it will remain there when we remove the string. However, one may worry that this process could destroy the wormhole. We have already seen that when the mass $m$ sits at the bottom of the wormhole, its energy cannot be larger than the bound \eqref{mbound} without collapsing the throat. We want to make sure that the energy that we are dropping into the wormhole as we lower the mass remains sufficiently small throughout the entire process.

We thus generalize our previous study to now hold the mass at an arbitrary height $\rho=\rho_0\geq 0$. Then the stress tensor of the (smeared) mass is
\begin{equation}
    T_{\tau\tau}^\delta=\frac{\beta}{4\pi G_N}\left( 1+\rho_0^2\right)\delta(\rho-\rho_0)\,,
\end{equation}
where $\beta$ is the same parameter for the mass $m$ as in \eqref{alphabeta}. The factor $(1+\rho_0^2)$ accounts for the fact that, since we keep fixed the black hole mass $m$ as measured locally at $\rho=\rho_0$, the energy conjugate to the time $\tau$ varies with the redshift along the wormhole tube.

To counterbalance the gravitational potential that pulls down on the mass, we employ a cosmic string whose tension pulls it upwards. Since the mass is smeared on $S^2$, the string will also have to be smeared, so the stress tensor is a step function of the form
\begin{equation}
   8\pi G_N \left(T_\tau{}^\tau\right)^{\rm string}=8\pi G_N \left(T_\rho{}^\rho\right)^{\rm string}=-\frac{T}{4\pi r_e^2}\Theta(\rho-\rho_0)\,.
\end{equation}

The string tension $T$ will be determined by solving the Einstein equations. The energy equation $(\tau\tau)$ now becomes
\begin{equation}
    (1+\rho^2)\phi''+\rho\phi'-\phi=\frac{\alpha}{1+\rho^2}-2\beta\,\delta(\rho-\rho_0)-\frac{T}{4\pi}\,\Theta(\rho-\rho_0)\,.
\end{equation}
When we require that the solution is continuous at $\rho=\rho_0$ we find that the tension must take the value
\begin{equation}\label{Tequil}
    T=\frac{8\pi \rho_0}{1+\rho_0^2}\beta\,.
\end{equation}
This tension remains finite all throughout the lowering process. It vanishes for $\rho_0\to\infty$ and $\rho_0\to 0$, which is as expected since these correspond to the beginning of the process and to the moment when we release the mass at its final equilibrium position. The solution for $\phi$ is readily found to be\footnote{See Appendix \ref{appendix:complete} for the complete solution to Einstein's equations.}
\begin{equation}\label{phisol2}
    \phi(\rho) = \alpha(1 + \rho \arctan\rho) - \frac{2\beta}{1+\rho_0^2}\left(\Theta(\rho-\rho_0)(\rho-\rho_0)-k \rho\right)\,.
\end{equation}
The integration constant $k$ depends on details of the physics of the lowering, and in particular on $\rho_0$, but we expect it to vary in the range
\begin{equation}\label{Crange}
    0\leq k\leq 1/2\,.
\end{equation}
When the mass is at $\rho_0\to\infty$, we expect to have $k=0$, which corresponds to just having an additional mass $m$ at one mouth and no strings, without any effect at the other mouth at $\rho\to-\infty$. Instead, when the mass reaches the bottom at $\rho=0$ we will have $k=1/2$, which is the fully symmetric solution that we obtained in \eqref{phisol}. 

Earlier we saw that the condition that the wormhole remains open is that $\phi$ grows for large $|\rho|$. In the solution \eqref{phisol2} this requires that
\begin{equation}\label{safeC}
    \frac{2\beta}{1+\rho_0^2}(1-k)<\frac{\pi}{2}\alpha\,.
\end{equation}
This bound becomes very weak when $\rho_0$ is large. When $\rho_0\to 0$, so that $k\to 1/2$, we recover the previous bound \eqref{mbound}. Without the detailed dependence of $k$ on $\rho_0$ we cannot know for certain whether a more stringent condition occurs at some finite $\rho_0\neq 0$. Nevertheless, it is clear that when
\begin{equation}
    \beta<C\frac{\pi}{2}\alpha\,,
\end{equation}
that is,
\begin{equation}\label{safe}
   m<C\frac{N_f Q}{8 r_e}=\frac{C}{8\sqrt{\pi}}g N_f m_P\,,
\end{equation}
with $C$ a number $\in [1/2,1]$, then we can safely lower the full mass $m$ to the bottom without collapsing the wormhole. 

Since the object being lowered can be a semiclassical black hole of mass $m$ if $N_f\gg 1$, this result also implies that information can be safely transmitted through the wormhole in single batches of the order of the black hole entropy $4\pi (m/m_P)^2$. It may be interesting to explore further how this type of analysis constrains the rates of information transfer through wormholes.

\subsection{Signaling between mouths}
\label{between}

Suppose that two parties, $A$ and $B$, use the original two-mouth wormhole to exchange messages. What are the consequences of inserting a third, small mouth operated by $c$? From the gravitational perspective, there are two different kinds of effects. First, the message sent by $A$ (a particle or a wave) may be partly absorbed by the small mouth and thus be received by $c$ and not $B$. The wormhole has then become a leaky pipeline. The absorption probability is proportional to the area of the small mouth, and thus to $c$, and can also have a dependence on the small mouth's angular position in the $S^2$ of the large throat. 

Due to the relation between traversing a wormhole and quantum teleportation, these effects will have counterparts in the entanglement structure of the three-mouth wormhole. In the absence of specific realizations it is difficult to be precise, but some qualitative features are plausibly realized. The leakiness of the line will likely appear as soon as a channel for a third party is added, with losses plausibly proportional to the number of degrees of freedom that $c$ holds. The angular dependence requires a more detailed understanding. In a wormhole where the throat geometry is well approximated by $AdS_2\times S^2$, any dual description will contain a sector modelled by quantum-mechanical degrees of freedom charged under an approximate $SO(3)$ (or $SU(2)$) symmetry group. Then, $A$ and $B$ can control the angular position on the sphere of the messages they exchange by selecting qubits with appropriately chosen charge distributions. Having information about this angular position is essential if $A$ and $B$ intend to communicate efficiently with $c$: the subset of degrees of freedom of their many-qubit system that hold the entanglement with $c$ must carry appropriate  $SO(3)$ charges.

A second effect is due to the Shapiro time delay that the signal will experience as it travels in the vicinity of the small mouth within the throat. That is, if the small mouth is placed in the throat geometry \eqref{eq:addphi}, then the signal that $A$ sends to be $B$ will take an additional time to arrive.  As measured by an observer in the throat, this delay is given by the familiar Shapiro formula 
\begin{equation}\label{shapiro}
    \delta t \sim 2m\log(\frac{4 \ell^2}{b^2}),
\end{equation}
where we have used the AdS$_2$ scale $\ell$ as an infrared cutoff and where $b$ is the distance of closest approach of the signal to the small mouth. This distance $b$ may be translated into an angular difference between the positions in $S^2$ of the mouth and the initial signal. However, the delay measured by $A$ and $B$ is much larger due to the strong redshift or order $\sim \ell/r_e$ at the bottom of the AdS$_2$ throat.
Here we might suppose that the increased travel-time can be correlated with an increased complexity in decoding the teleported message.


\section{Discussion}
\label{sec:disc}

In the above work, we presented a construction of a multi-mouth traversable wormhole. This was done by starting with a two-mouth traversable wormhole of the form described in \cite{Maldacena:2018gjk}. We then perturbed this solution by adding a small black hole in the throat of the larger wormhole. So long as the total energy remains below the energy of a pair of disconnected extremal black holes,  the original wormhole will remain traversable. With large enough numbers $N_f$ of bulk fermion fields, this can be accomplished for small black holes far bigger than the Planck length $l_P$, and even also bigger than the effective cutoff length $\sqrt{N_f}l_P$, so that the term `black hole' is appropriate. This small black hole can be placed in mechanical equilibrium through the proper placement of cosmic strings. Additional compact cosmic strings can be used to make all mouths traversable.

To maintain traversability, the small black hole in the throat must have an energy below the gap $E_{gap}$ described in \cite{Maldacena:2018gjk} that separates traversable wormholes from black holes.  It achieves this in part due to the strong redshift in the throat, so that the black hole's energy is {\it much} smaller than that of a similarly-sized black hole in the region outside the wormhole.  Nevertheless, we saw in Sec.~\ref{subsec:lowering} that such a black hole can be (carefully) lowered into the wormhole throat from outside without causing the wormhole to collapse.  In particular, it shows that there can be configurations in which the total mass contained within the wormhole is quite far above the energy associated with separate black holes and in which the wormhole remains traversable, so long as this energy is not localized too deep within the wormhole throat.

A rather different construction of a traversable three-mouth wormhole was recently described in \cite{Balushi:2020lyc}.  That analysis began with a non-traversable three-boundary wormhole asymptotic to AdS$_3$ and added boundary interactions similar to those in the original work by Gao, Jafferis, and Wall \cite{Gao:2016bin}.   In particular, by taking a limit where the horizons that shroud the original non-traversable wormhole become both very large and very hot, much as in \cite{Marolf:2015vma}, the causal shadow becomes exponentially thin along large regions of the horizon.  In fact, in such regions the wormhole geometry becomes exponentially close to that of the BTZ version of the Einstein-Rosen bridge.  It is thus straightforward to apply a local version of the analysis of \cite{Gao:2016bin} to show that appropriate boundary interactions can make the wormhole  traversable between any two boundaries.

The fact that the analysis of \cite{Balushi:2020lyc} largely reduces to that of \cite{Gao:2016bin} is associated with the fact that the entanglement structure of the non-traversable three-boundary wormhole reduces in this this limit (and in the relevant regions of the boundary) to the entanglement structure of the thermofield double.  To be specific, in the region where the causal shadow separating boundaries $A$ and $B$ becomes very thin, the corresponding parts of the dual field theory on boundaries $A$ and $B$ are exponentially well approximated by a thermofield double state  \cite{Marolf:2015vma}. In particular, neglecting exponentially small corrections we may say that this part of boundary $A$ is entangled {\it only} with boundary $B$ and has no entanglement with $C$.  In this sense, the traversability of the three-mouth wormhole constructed in \cite{Balushi:2020lyc} is associated only with bipartite entanglement; thus, multipartite entanglement plays no role.

In contrast, multiparty entanglement seems likely to play an important role in the three-mouth traversable wormhole constructed in the current work. To make the discussion precise, we consider an asymptotically AdS version of our construction in which each of the three mouths is associated with a separate asymptotic region (the negative energy then comes not from fermion loops but from operator insertions at the mouths as in \cite{Gao:2016bin,Maldacena:2018lmt,Balushi:2020lyc}). Our three-mouth wormhole then becomes a three-boundary wormhole.  In the limit where the AdS scale $\ell$ is large compared with the radii of the throats, the local geometry of the throats will be identical to that of the asymptotically flat case described in the main text.

\begin{figure}[h]
        \begin{center}
         \includegraphics[width=0.45\textwidth]{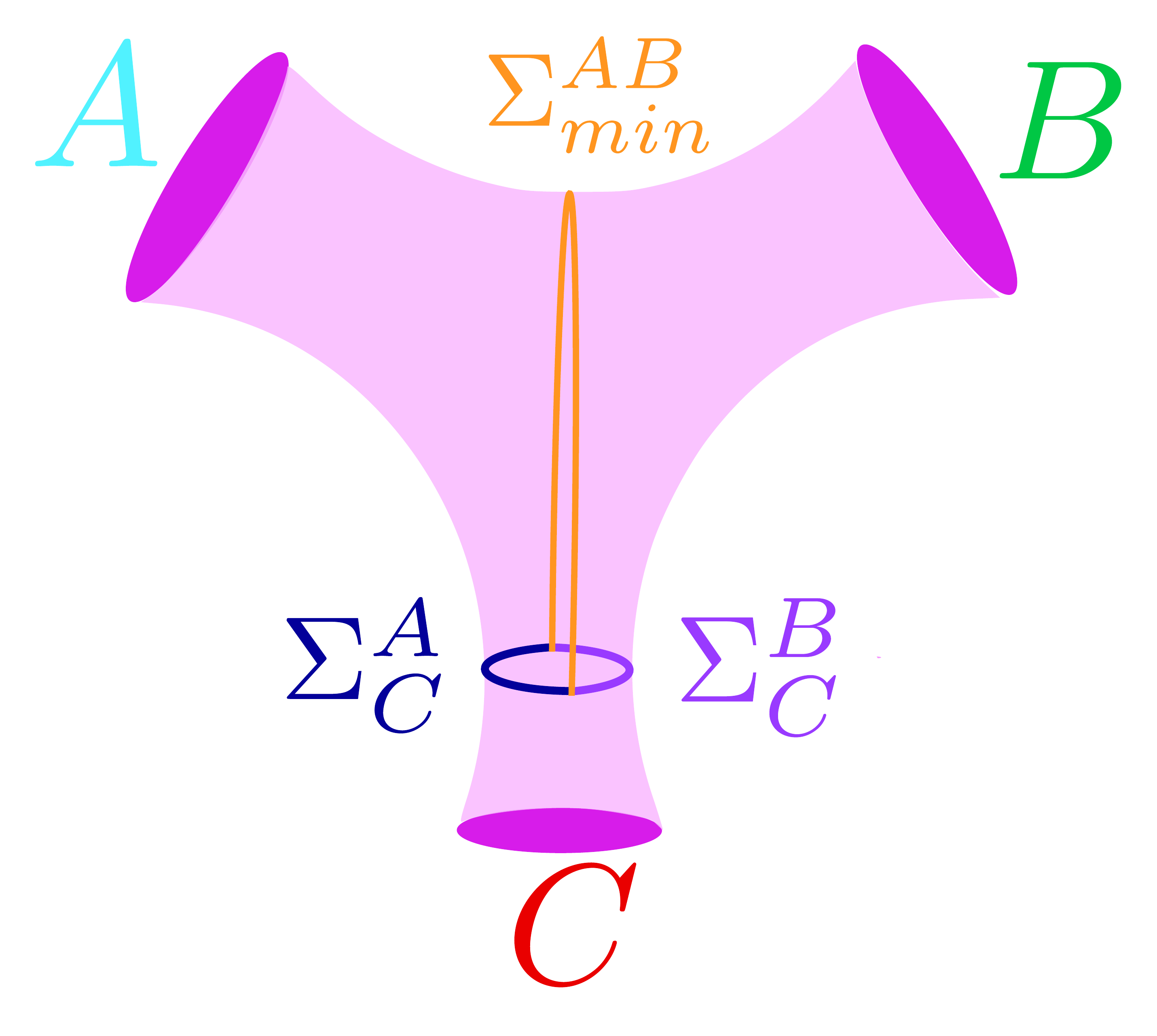}
        \end{center}
        \caption{\small A three-mouth wormhole. $\Sigma_{min}^{AB}$ is the minimal surface homologous to both $A$ and $B$ that stays within the entanglement wedge of $AB$ (orange). $\Sigma_C^A$ (dark blue) and $\Sigma_C^A$ (purple) are the portions of $\Sigma_C$, split by $\Sigma_{min}^{AB}$, such that $\Sigma_{min}^{AB}\cup\Sigma_C^A$ is homologous to $A$ and $\Sigma_{min}^{AB}\cup\Sigma_C^B$ is homologous to $B$.}
        \label{fig:ewcs}
\end{figure}

To argue for the possible importance of multiparty entanglement in our case, let us first recall from \cite{Akers:2019gcv} that multiparty entanglement may be quantified by considering the entanglement wedge $W_{AB}$ of the joint $AB$ system and computing 
\begin{equation}
M_3: = 2E_W(AB) - I(A:B),
\end{equation}
where $E_W(AB)$ 
is the entanglement wedge cross section entropy \cite{Umemoto_2018} and $I(A:B)$ is the mutual information between $A$ and $B$.  In particular, $E_W(AB)$ is $1/4G$ times the area $A(\Sigma^{AB}_{min})$ of the minimal surface homologous to both $A$ and $B$ within the entanglement wedge, where the homology condition now allows the homology surfaces to have additional boundaries at finite boundaries of $W_{AB}$, the entanglement wedge of $AB$. In particular, the surface $\Sigma^{AB}_{min}$ will generally intersect the minimal surface $\Sigma_C$ that computes the entropy of boundary $C$, and indeed will split it into two parts $\Sigma_C^A$ and $\Sigma_C^B$.  As a result, since $S_{AB}=S_C$ for our wormhole, we may rewrite $M_3$ in the form
\begin{equation}
\label{eq:M3pos}
    4G M_3 = \left( A(\Sigma^{AB}_{min})  + A(\Sigma_C^A) - A(\Sigma_A)\right) + \left( A(\Sigma^{AB}_{min})  + A(\Sigma_C^B) - A(\Sigma_B)\right), 
\end{equation}
where $\Sigma_A, \Sigma_B$ are the minimal surfaces homologous to boundaries $A$ and $B$ in the usual sense.
The right hand side of \eqref{eq:M3pos} is now manifestly positive since, for example, $\Sigma_A$ is homologous to $\Sigma^{AB}_{min}\hspace{1pt} \cup \hspace{1pt} \Sigma_C^A$ and is also by definition minimal within that homology class; see Fig.~\ref{fig:ewcs}.  
Furthermore, in the limit used in the main text in which the AdS scale $\ell$ and the radii of the large mouths are much larger than the radius of the third small mouth, it is clear that the third small mouth sets the only scale in the problem.  Thus in that case, dimensional analysis guarantees that $4GM_3$ will be first order in the area of the third small mouth, or in other words, that $M_3$ is first order in the corresponding entropy: $M_3 \sim S_C$. 

This shows that our construction applies in the limit where the multi-mouth wormhole has significant multiparty entanglement. The remaining question is thus whether this multiparty entanglement plays an important role in our wormhole's traversability.  While a complete analysis of this question is beyond the scope of the current work, the further remarks below appear to point in this direction.

Let us briefly consider the locations of $\Sigma_A, \Sigma_B$, and $\Sigma_C$. In \cite{Marolf:2015vma,Balasubramanian:2014hda}, it was found that narrowing of the entanglement shadow region between these three surfaces, so that the separation between some two of these surfaces becomes small relative to their distance to the third, was indicative of a region of mostly bipartite entanglement between the corresponding boundaries. In contrast, regions where the distance between the various entangling surfaces is roughly the same between each pair of surfaces might naturally be taken as a signal of tripartite entanglement. In particular, \cite{Balasubramanian:2014hda} associated large amounts of multipartite entanglement to certain AdS black holes whose temperature was small  compared to the AdS scale, while \cite{Marolf:2015vma} showed that states dual to hot black holes  are well-approximated by sewing together various copies of $|TFD\rangle$ states. See Fig.~\ref{fig:entshadow} below.

\begin{figure}[h]
        \begin{center}
        \begin{subfigure}{0.45\linewidth}
        \centering
         \includegraphics[width=0.7\textwidth]{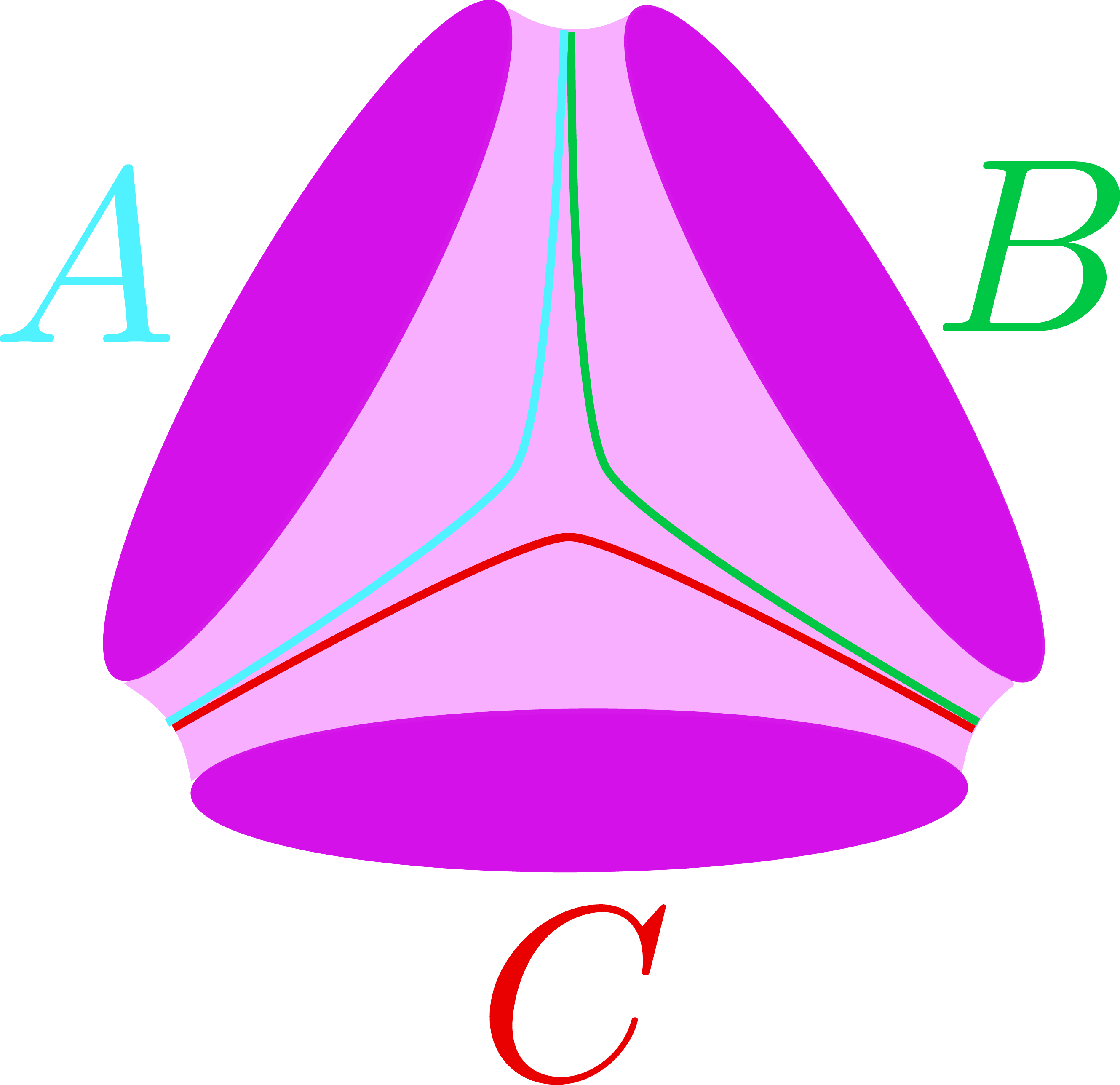}\quad
                \end{subfigure}
        \begin{subfigure}{0.45\linewidth}
        \centering
         \includegraphics[width=0.8\textwidth]{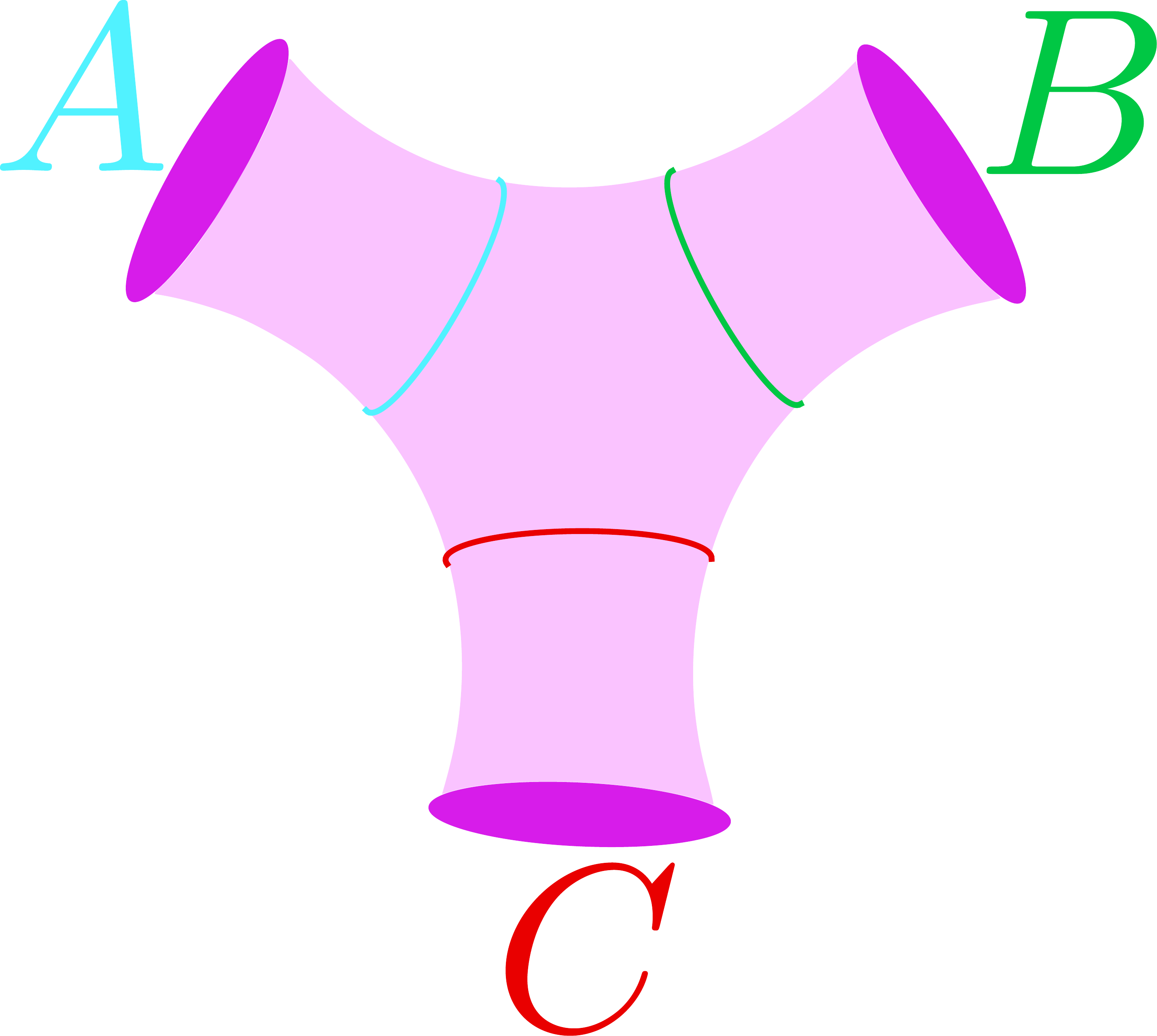}\quad
                \end{subfigure}
        \end{center}
        \caption{\small {\bf{Left}}: A hot AdS$_3$ three-mouth wormhole.  The entanglement shadow becomes very narrow in regions where pairs of extremal surfaces approach each other, indicating regions of strong bipartite entanglement. {\bf{Right}}: A cold AdS$_3$ three-mouth wormhole. At any point on one entangling surface, the distance to the other two entangling surfaces is roughly the same.  This suggests strongly tripartite entanglement. }
        \label{fig:entshadow}
\end{figure}

Recall also that our analysis of back-reaction suggested that one mouth ($C$) must remain small relative to the other two.  We thus assume that this is so.  Before we add in $C$, the extremal surfaces associated with $A$ and $B$ coincide and lie at the bottom of the $AB$ throat. In the limit where $C$ is much smaller than $A$ and $B$, it will have little effect on the geometry far from $C$.  Thus the extremal surfaces associated with $A$ and $B$ will remain close over most of their area, and in particular at the top of the wormhole in Fig.~\ref{fig:entshadow2} below.  
This indicates that there is still strong two-party entanglement between $A$ and $B$, as one would expect since $S_C \ll S_A,S_B$.


On the other hand,  the extremal surface associated with $C$ will remain close to the bottom of the small wormhole throat.   Since the only scale in the problem in the region near $C$ is the size of $\Sigma_C$, the separation in this region between any two minimal surfaces will be comparable to the scale of $\Sigma_C$ itself.   The fact that e.g. the separation in this region between $\Sigma_A$ and $\Sigma_B$ is comparable to the separation between $\Sigma_A$ and $\Sigma_C$ then indicates that multiparty entanglement plays an important role in this region of the geometry, and thus presumably also in making this region traversable.  More physically, one might rephrase this remark by stating that a signal entering the wormhole through mouth $C$ must then find itself for some non-trivial amount of time in the entanglement shadow region which, due to the entanglement with $C$, fails to be part of either of the entanglement wedges of $A$ or $B$ alone.  One thus expects that the three-party entanglement of the field theory dual is required to describe propagation of the signal in this region.

\begin{figure}[h]
        \begin{center}
        \begin{subfigure}{0.48\linewidth}
        \centering
         \includegraphics[width=0.9\textwidth]{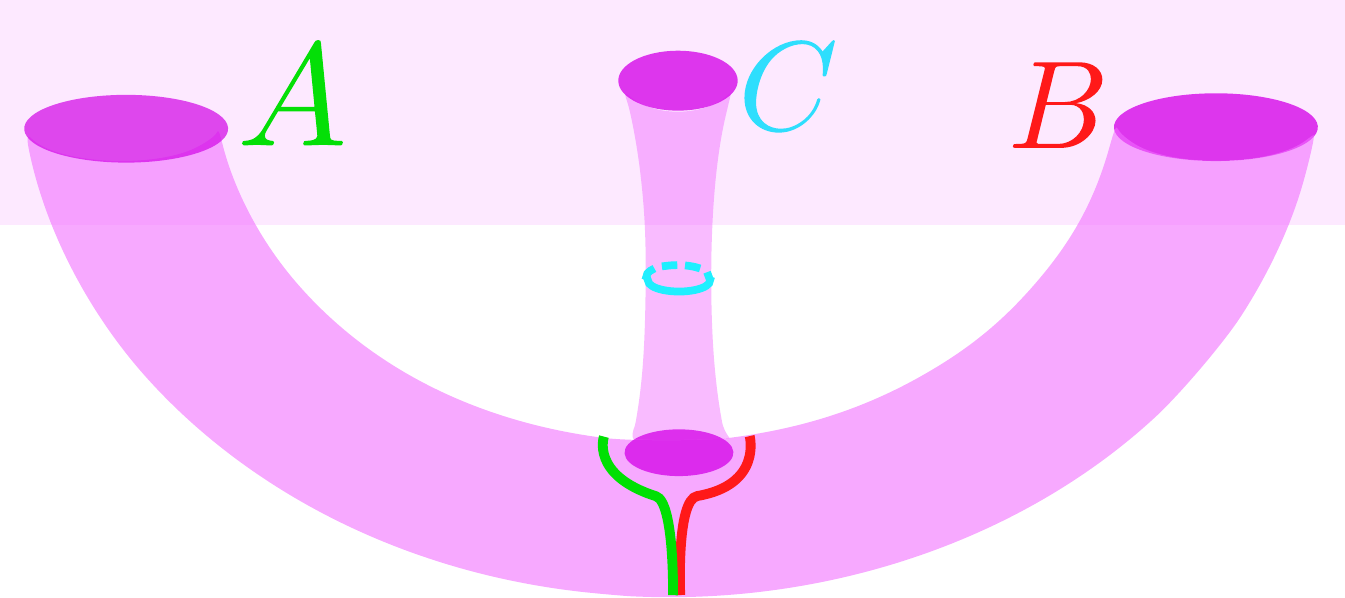}\quad
                \end{subfigure}
        \begin{subfigure}{0.48\linewidth}
        \centering
                \end{subfigure}
        \end{center}
        \caption{\small The extremal surfaces associated with the mouths of a multi-mouth wormhole in asymptotically flat space are depicted, with one mouth $C$ much smaller than the other two, $A$ and $B$. In this limit, the extremal surfaces associated with $A$ and $B$ remain close together   over most of their area, and in particular much closer than their distance to the extremal surface associated with $C$, which sits at the bottom of the small wormhole throat.}
        \label{fig:entshadow2}
\end{figure}


The large bipartite entanglement between $A$ and $B$ is consistent with the idea that $C$ has little effect on signals being sent between $A$ and $B$. But it would be interesting to consider quantum mechanical duals in more detail, as well as the quantum teleportation protocols associated with traversing the wormhole in the bulk in analogy with the discussions of e.g. \cite{Gao:2016bin,Maldacena:2017axo,Freivogel:2019whb,Berenstein:2019yfv}.
In particular, if the entanglement of $C$ with $A$ and $B$ is indeed mostly of the multi-party sort, then the dual description of sending a signal from $C$ to either one of $A$ or $B$ must necessarily involve all three systems.  While this idea may at first seem unfamiliar, it is consistent with the fact that the asymptotically flat region of our gravitational solution does in fact provide interactions between each pair of mouths $AB$, $AC$, and $BC$.

\section*{Acknowledgments}
We thank Chris Akers and Sergio Hern\'{a}ndez-Cuenca for useful discussions.  RE and MT were supported by ERC Advanced Grant GravBHs-692951, MEC grant FPA2016-76005-C2-2-P, and AGAUR grant 2017-SGR 754.  BG-W was supported by a University of California, Santa Barbara central fellowship. DM was supported in part by NSF grant PHY1801805 and by funds from the University of California.  MT was also supported in part by the Heising-Simons Foundation, the Simons Foundation, and National Science Foundation Grant No. NSF PHY-1748958. This work was partly carried out while RE, DM, and MT were participants in a KITP program, where their research was supported in part by
the National Science Foundation under Grant No. NSF PHY-1748958 to the KITP.

\appendix


\section{Matched asymptotics construction}
\label{appendix:mae}

At a technical level, the insertion of the small mouth in the initial wormhole throat can be carried out as a typical construction of matched asymptotics. The throat characteristic size $r_e$ is much larger than the small mouth radius, $G_N m$, so we consider a ``near zone'' around the small mouth where $r\ll r_e$, and a ``far zone'' in the throat where $r\gg G_N m$. In the far zone the small mouth is well approximated as a pointlike source whose effect can be incorporated as a small perturbation of the throat. In the near zone, the mouth is approximated as a near-extremal RN black hole, which is not exactly round nor asymptotically flat, since it is tidally distorted by the curvature of the throat. The two zones overlap where $G_N m\ll r \ll r_e$, and this allows to match the geometries and the parameters in them, so as to build a global solution for the throat (which then is itself matched to the exterior at $\rho\gg 1$, i.e., at distances $\gg r_e$ in the throat). 

The construction is thus reduced to solving two different problems of gravitational perturbations, in the near and far zones. In the following, we describe how these problems are set up in the different zones, and how they allow us to obtain the balance of forces inside the throat. The explicit backreacted solution in the region near the throat exit at $\rho\gg 1$ is explained in  Sec.~\ref{Malda}.

\subsection*{Near zone}

We want to insert a small RN black hole in the AdS$_2\times S^2$ throat geometry
\begin{equation}
\label{ads2s2}
    ds^2=r_e^2\left( -(1+\rho^2)d\tau^2+\frac{d\rho^2}{1+\rho^2}+d\Omega^2\right)\,,
\end{equation}
which is threaded with magnetic field lines with gauge potential
\begin{equation}\label{Fthroat}
A_\phi=\frac{Q}{2}\left(1-\cos\theta\right)\,.
\end{equation}
For our purposes here, we neglect the small corrections to the geometry $\gamma$ and $\phi$ \eqref{eq:addphi}. We will also consider that the insertion is an extremal RN black hole, neglecting that it is slightly above extremality. Both effects can be well controlled parametrically and can be added if desired by slightly perturbing the construction below.

The near zone is the geometry within a region of size $\ll r_e$ around a point at some value $\rho=\rho_0$ along the throat and at an arbitrary position in the $S^2$. We allow for a finite displacement $\rho_0\neq 0$ away from the throat bottom. As we will see, this plays an important role when all the mouths are charged under the same $U(1)$: the resulting gravitational pull towards the bottom can compensate for the magnetic force that the field \eqref{Fthroat} exerts on the small mouth, and stabilize its position.\footnote{This displacement is a different effect than discussed in Sec.~\ref{subsec:lowering}. Both can be treated separately, and combined if desired.}

We introduce near-zone cylindrical coordinates $(\bar{t},\bar{\bf{x}})=(\bar{t},\bar{z}, \bar{\sigma},\phi)$, aligned with the throat and centered on the chosen point, such that
\begin{equation}
    \bar{t}=r_e\sqrt{1+\rho_0^2}\,\tau\,,\qquad \bar{z}=\frac{r_e}{\sqrt{1+\rho_0^2}}\left(\rho-\rho_0\right)\,,
\end{equation}
and $\bar{\sigma}$ is a stereographical polar coordinate on the $S^2$ of radius $r_e$,
\begin{equation}
     \bar{\sigma} = 2r_e \tan \frac{\theta}{2} \,,
\end{equation}
so that
\begin{equation}
    r_e^2d\Omega^2=\frac{d\bar{\sigma}^2+\bar{\sigma}^2 d\phi^2}{\left(1+\frac{\bar{\sigma}^2}{4r_e^2}\right)^2}\,.
\end{equation}

Then, when $|\bar{\bf{x}}|\ll r_e$, \eqref{ads2s2} becomes
\begin{align}\label{nearasym}
    ds^2\simeq & -\left( 1+2A \bar{z}+\frac{\bar{z}^2}{r_e^2}\right)d\bar{t}^2
    +\left( 1-2A \bar{z}+\left(1-4A^2 r_e^2\right)\frac{\bar{z}^2}{r_e^2}\right)d\bar{z}^2\nonumber\\
    &+\left(1-\frac{\bar{\sigma}^2}{2r_e^2}\right)\left(d\bar{\sigma}^2+\bar{\sigma}^2 d\phi^2\right)+\order{\frac{\bar{\bf{x}}}{r_e}}^3\,,
\end{align}
and the magnetic potential 
\begin{equation}\label{asymF}
A_\phi=\frac{Q}{4}\frac{\bar{\sigma}^2}{r_e^2}\left(1-\frac{\bar{\sigma}^2}{4r_e^2}
+\order{\frac{\bar{\bf{x}}}{r_e}}^3\right) \,.
\end{equation}
We have introduced a parameter
\begin{equation}
    A=\frac{\rho_0}{r_e\sqrt{1+\rho_0^2}}\,,
\end{equation}
which is to be interpreted as an acceleration, since the terms linear in $\bar{z}$ in the metric correspond to the gravitational potential that accelerates a particle at $\rho=\rho_0$ towards the throat bottom at $\rho=0$. The almost uniform magnetic field along the $\bar{z}$ direction is
\begin{equation}\label{Bfield}
    B=\frac{Q}{2r_e^2}\,.
\end{equation}

In the limit where $r_e\to\infty$ keeping  $\bar{\bf{x}}$ finite, we recover Minkowski space in \eqref{nearasym}. The corrections for small $\bar{\bf{x}}/r_e$ account for the modification of the near-zone due to the curvature of the throat. They give rise to tidal distortions on any object localized within this zone. More concretely, when we insert an extremal RN black hole, the metric will take the form
\begin{align}\label{metpert}
    ds^2\simeq & -\left( 1+2A \bar{z}+\frac{\bar{z}^2}{r_e^2}\right)H^{-2}(\bar{\bf{x}})d\bar{t}^2
    +\left( 1-2A \bar{z}+\left(1-4A^2 r_e^2\right)\frac{\bar{z}^2}{r_e^2}\right)H^{2}(\bar{\bf{x}})d\bar{z}^2\nonumber\\
    &+\left(1-\frac{\bar{\sigma}^2}{2r_e^2}\right)H^{2}(\bar{\bf{x}})\left(d\bar{\sigma}^2+\bar{\sigma}^2 d\phi^2\right)\nonumber\\
    &+\left(\frac{G_N m}{r_e}h^{(1)}_{\mu\nu}(\bar{\bf{x}})+\frac{(G_N m)^2}{r_e^2}h^{(2)}_{\mu\nu}(\bar{\bf{x}})\right)d\bar{x}^\mu d\bar{x}^\nu+\order{\frac{\bar{\bf{x}}}{r_e}}^3\,,
\end{align}
and the magnetic potential
\begin{equation}\label{Fpert}
A_\phi=\frac{Q}{4}\frac{\bar{\sigma}^2}{r_e^2}\left(1-\frac{\bar{\sigma}^2}{4r_e^2}\right)
+\frac{q}{2}\left(1-\frac{\bar{z}}{\sqrt{\bar{z}^2+\bar{\sigma}^2}}\right)
+\frac{G_N m}{r_e}a^{(1)}_\phi+\frac{(G_N m)^2}{r_e^2}a^{(2)}_\phi+\order{\frac{\bar{\bf{x}}}{r_e}}^3\,.
\end{equation}
Here
\begin{equation}\label{Hx}
    H(\bar{\bf{x}})=1+\frac{G_N m}{|\bar{\bf{x}}|}=1+\frac{G_N m}{\sqrt{\bar{z}^2+\bar{\sigma}^2}}
\end{equation}
has been introduced such that, when  $r_e\to\infty$ keeping  $\bar{\bf{x}}$ and $m$ finite, we recover the extremal RN black hole of mass $m$ and (integer) charge
\begin{equation}
    q=\frac{g l_P}{\sqrt{\pi}}m
\end{equation}
in isotropic cylindrical coordinates. When $r_e$ is large but finite, the functions $h^{(1,2)}_{\mu\nu}$ and $a^{(1,2)}_\phi$ account for the small distortion and polarization that the tidal field and the magnetic field induce on the black hole. These functions are determined by solving the Einstein equations for a perturbation of the RN black hole, with the condition that the solution remains regular at the black hole horizon, and that it asymptotes to \eqref{nearasym} when $|\bar{\bf{x}}| \gg G_N m$. 

The tidal fields in the asymptotic region are of two kinds: the leading effect is the linear, dipole field $\sim A\bar{z}$ that, as we mentioned, acts as an accelerating gravitational potential. We will presently obtain the solution for the corresponding perturbation $h^{(1)}_{\mu\nu}$, and in this way fix $A$ so that the gravitational pull towards the bottom exactly balances the magnetic force of the background field \eqref{asymF} towards the oppositely charged mouth.

The subleading terms, quadratic in $\bar{\bf{x}}/r_e$, represent a quadrupolar tidal field. When $\rho_0=0$, i.e., $A=0$, the corresponding perturbations $h^{(2)}_{\mu\nu}$ also appear in the calculation of the Love numbers of a black hole as a measure of its deformability in linear response theory. This problem has been solved for RN black holes in \cite{Cardoso:2017cfl}, where the explicit results for the metric perturbation can be found. When $A\neq 0$ the perturbations $h^{(2)}_{\mu\nu}$ depend on the solution for $h^{(1)}_{\mu\nu}$ and the problem becomes more complicated. Since we do not need them, we will not pursue their calculation further in this article.

\subsubsection*{Force balance}

The following analysis is, as we have indicated, relevant only when there is an interaction between the magnetic charge of the small mouth and the magnetic field in the throat. In models with several $U(1)$'s these forces can be avoided and the small mouth then sits at equilibrium at $\rho=\rho_0=0$ with $A=0$. However, when all the magnetic charges are in the same gauge field, we expect to be able to equilibrate their forces against the gravitation in the throat.

In order to simplify the analysis, we can use the equivalence principle in the region $|\bar{\bf{x}}|\ll A^{-1}$ around the small mouth to trade the tidal gravitational field for an acceleration. This transforms the problem into that of a magnetic black hole uniformly accelerated by an external magnetic field. An exact solution in the Einstein-Maxwell theory for this system was given by Ernst in \cite{Ernst:1976}. It combines the features of the C-metric for an accelerating black hole with those of a magnetic Melvin flux tube. The solution has been much further studied following  \cite{Garfinkle:1990eq}, so we will be brief. 

We are interested in the particular case of a RN black hole of mass $m$ and charge $q$, immersed in a magnetic field $B$ and moving with acceleration $A$. This will be an adequate approximation to our system when the black hole is small, in the sense that $G_N m, q l_P \ll (B l_P)^{-1}, A^{-1}$, and within distances away from the black hole smaller than $A^{-1}$. This region excludes the acceleration horizon of the Ernst solution, which is not present in the situation of interest to us. One can easily show that the horizon of the black hole is extremal (degenerate) when $q =\frac{gl_P}{\sqrt{\pi}}m\left(1+\order{G_N mA}^2\right)$. 

In this limit, the absence of conical singularities at the axis of rotational symmetry is equivalent to the requirement that Newton's law is obeyed \cite{Ernst:1976}, 
\begin{equation}
    qB\simeq mA(1+\order{mA}^2)\,.
\end{equation}
For an extremal black hole this is
\begin{equation}
    B=\frac{\sqrt{\pi}}{g l_P}A(1+\order{mA}^2)\,.
\end{equation}
If we use that the magnetic field $B$ along the wormhole throat is given by \eqref{Bfield},
then we find that the equilibrium position, $\rho_0$, of the small mouth in the wormhole is given by
\begin{equation}
    \frac{\rho_0}{\sqrt{1+\rho_0^2}}=\frac{gl_P}{\sqrt{\pi}}\frac{Q}{2r_e}=\frac{g^2}{2\pi}\,,
\end{equation}
that is,
\begin{equation}
    \rho_0=\frac{g^2}{2\pi}\frac1{\sqrt{1-\frac{g^4}{4\pi^2}}}\,.
\end{equation}
For fixed $g$, $\rho_0$ is independent of the wormhole charge. The physical distance $r_e \rho_0$ along the wormhole grows linearly with $Q$, as expected.

It is now possible to find the near-zone geometry with the leading order backreaction from the tidal  acceleration. To do so, we expand the Ernst solution within distances $\ll A^{-1}$ of the black hole, keeping only terms up to linear order in $A$. Doing this, we obtain the metric\footnote{The usual C-metric $(x,y)$ coordinates are transformed into $(\bar{z},\bar{\sigma})$ as $y^{-1}=A(\sqrt{\bar{z}^2+\bar{\sigma}^2}+m)$ and $x=\frac{\bar{z}}{\sqrt{\bar{z}^2+\bar{\sigma}^2}}-mA/2$.}
\begin{align}
    ds^2\simeq &-\left(1+2A \bar{z}\right)H^{-2}(\bar{\bf{x}})d\bar{t}^2
    +\left(1-2A \bar{z}\right)H^{2}(\bar{\bf{x}})d\bar{z}^2\nonumber\\
    &+\left(1-2G_N m A\frac{\bar{z}}{\sqrt{\bar{z}^2+\bar{\sigma}^2}}\right)H^2(\bar{\bf{x}}) \left(d\bar{\sigma}^2+\bar{\sigma}^2 d\phi^2\right)
\end{align}
where $H(\bar{\bf{x}})$ is the same as in \eqref{Hx}
and the magnetic potential is
\begin{equation}
    A_{\phi}=\frac{q}{2}\left(1-\frac{\bar{z}}{\sqrt{\bar{z}^2+\bar{\sigma}^2}}\right) +\frac{B}{2}\left(\bar{\sigma}^2-(G_N m)^2\left(1+\frac{\bar{z}^2}{\bar{z}^2+\bar{\sigma}^2}\right)\right)\,.
\end{equation}
These give the perturbations $h^{(1)}_{\mu\nu}$ and $a^{(1)}$ in \eqref{metpert} and \eqref{Fpert}.

\subsection*{Far and very far zones}

Having explained in detail the set up for the near-zone construction, our description of the far-zone analysis will be much more succinct. The far zone geometry is \eqref{ads2s2}, or more precisely, \eqref{eq:addphi}. Here the small near-extremal RN solution is inserted as a localized delta-function source of stress-energy and magnetic charge. These create a small distortion of the throat geometry \eqref{eq:addphi}, which must be solved as a perturbation of the Einstein equations. It is clear that a solution to the problem exists, and although its explicit form is likely to be complicated, fortunately we do not need it. 

We can obtain useful information more easily if we restrict ourselves to the asymptotic region of the far zone, where $\rho\gg 1$ -- the `very far zone'. Recall that the throat is measured in units of $r_e$, so these are distances $\gg r_e$, where the throat joins the exterior of the original wormhole. At these large distances from the small mouth, we cannot resolve the precise location in $S^2$ of the delta-source. Then, we can appropriately smear its position over the entire $S^2$, and regard it as a codimension-one defect along the $\rho$ direction (notice  that having $\rho\gg \rho_0$ also allows us to neglect $\rho_0$).

In this manner, we can obtain the effect that the insertion of the small mouth has on the junction of the throat to the exterior mouth. This analysis is presented in Sec.~\ref{Malda}, where we use it to determine how big $m$ can be before making such a junction impossible.

\section{Mass at a height in the throat}
\label{appendix:complete}

In Sec.~\ref{subsec:lowering} we presented the solution for the $\phi$ perturbation when we have a mass $m$ (smeared on the $S^2$) at a height $\rho=\rho_0$ in the throat of the wormhole, suspended from a cosmic string. Here we give the solution to the entire set of Einstein's equations. The $(\rho\rho)$ equation
\begin{equation}
    \rho\phi'-\phi=-\frac{\alpha}{1+\rho^2}-\frac{T}{4\pi}\,\Theta(\rho-\rho_0)\,,
\end{equation}
is automatically satisfied by \eqref{phisol2} with tension $T$ given by \eqref{Tequil}. The remaining equation, along the sphere directions, is
\begin{equation}
    \gamma''+(1+\rho^2)\phi''+2\rho \phi'+4\phi= 0\,,
\end{equation}
which is solved by
\begin{align}
    \gamma=&-\alpha\left((\rho^2+3)\rho\arctan\rho +\rho^2-\ln(1+\rho^2)\right)\nonumber\\
    &+\frac{\beta}{1+\rho_0^2}\left(
    |\rho-\rho_0|(1+\rho^2)+2\rho_0(\rho-\rho_0)^2\Theta(\rho_0-\rho)-3\rho_0\rho^2+(1-2k)\rho^3
    \right)\,.
\end{align}
The condition \eqref{safeC} ensures that $\gamma$ decreases at large $|\rho|$. This criterion could also have been taken as the condition for being able to match the wormhole to the exterior mouth and thus keep it open.

If we set $\rho_0=0$ and $k=1/2$ we obtain the solution for $\gamma$ for the configurations in Sec.~\ref{Malda}.

\section{Multi-mouth wormholes from a perturbed bifurcate horizon}
\label{appendix:pertwhs}

We can also consider using the wormholes of \cite{Fu:2018oaq,Fu:2019vco,Marolf:2019ojx}, which perturb around a bifurcate Killing horizon. These wormholes will generally be strongly time-dependent, and can be traversed by curves only if they leave past null infinity at sufficiently early times. This fact leads to a more stringent bound on the size of the smaller wormhole mouth.


\begin{figure}[h]
        \begin{center}
        \begin{subfigure}{0.45\linewidth}
        \centering
                \includegraphics[width=0.95\textwidth]{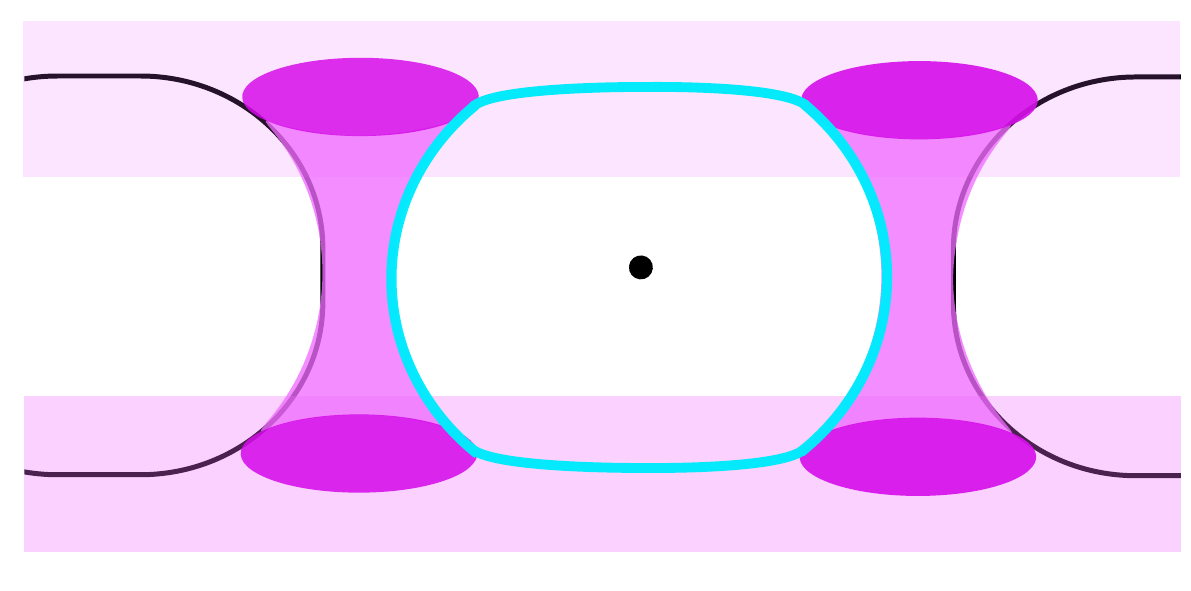}\quad
                \end{subfigure}
		\begin{subfigure}{0.4\linewidth}
        \centering
                \includegraphics[width=0.95\textwidth]{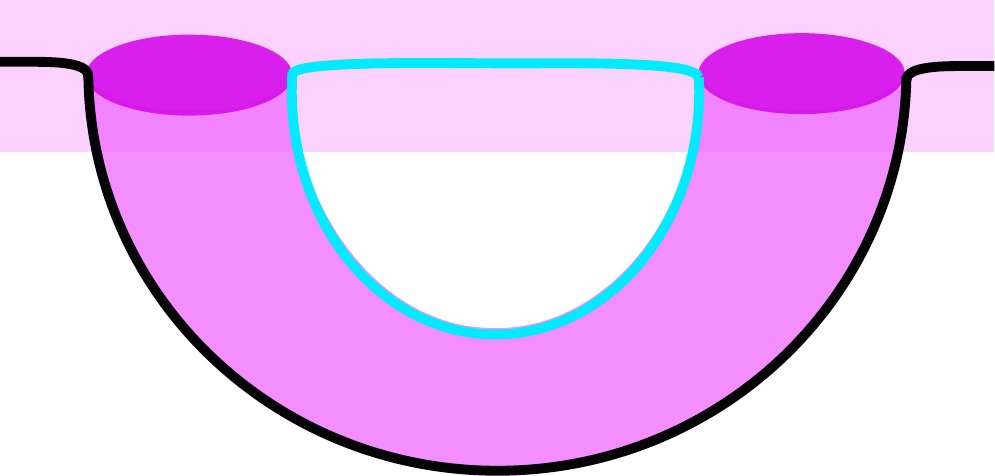}\quad
                \end{subfigure}
        \end{center}
        \caption{\small {\bf{Left}}: The quotient that creates a wormhole similar to Fig. \ref{fig:4d_mag} starts with two maximally extended black holes, held apart by cosmic strings that run off to infinity (black). A point is identified with the point given by swapping the black hole mouths and the asymptotic regions, e.g. it can be thought of as identifying a point with its $\pi$ rotation about the black dot in the figure. {\bf{Right}}: The result of this quotient gives a wormhole with two mouths in the same asymptotic region. Quantum fluctuations of the light blue compact string will generate negative energy.}
        \label{fig:quotient}
\end{figure}

This construction starts with a classical background containing a pair of maximally-extended, charged RN-like black holes, as in Fig.~\ref{fig:quotient}. These black holes are held apart by cosmic strings that run off to infinity. An additional compact cosmic string wraps the horizon -- and will be used to generate the Casimir energy needed for traversability. We can then quotient this background geometry by identifying a point with the one obtained by swapping the black hole mouths and the asymptotic regions (see Fig.~\ref{fig:quotient}).  This quotienting results in a wormhole with two mouths in the same asymptotic region, with the compact cosmic string becoming a shorter compact cosmic string, and the cosmic strings stretching to infinity becoming a single string that goes through the wormhole mouth and stretches to infinity on either side. 

Since the cosmic strings lie along the horizon, the classical cosmic string stress tensor, which is proportional to the induced metric, will not contribute to $\langle T_{kk}\rangle_M$. Quantum fluctuations of the string will contribute, however. In \cite{Fu:2019vco}, these fluctuations were modelled as 1+1 dimensional massless free scalar fields. 
For the string stretching to infinity, the points identified in the quotient will lie on two distinct non-compact strings in the covering space. The fluctuations on two different strings will be uncorrelated, and so the quantum fluctuations of the non-compact string will not contribute to $\int \langle T_{kk}\rangle_M d\lambda$. However, the contributions from quantum fluctuations of the compact string are non-zero, and give rise to the Casimir energy needed for traversability. 


In the extremal limit of  the RN black holes of the classical backgrounds, the back-reaction becomes large, and so the wormhole remains open for longer times; however, our ability to treat the back-reaction perturbatively will break down. At finite temperature, the fact that the wormholes are open for a finite amount of time limits the size of the black hole that can be placed in the throat. We compute that effect below. 

We note that from \cite{Fu:2019vco},
\begin{equation}
\Delta V = \frac{r_+}{r_-}{{\left( \frac{r_+-r_-}{r_-} \right)}^{-1-{{\left( {{r}_{-}}/{{r}_{+}} \right)}^{2}}}}{{e}^{2{{\kappa }_{+}}r_+}} \int_{-\infty}^{+\infty} \text{d}U h_{kk},
\end{equation}
where
\begin{equation}
\left( \int dU h_{kk} \right) (\Omega) = 8 \pi G \int d \Omega' H(\Omega, \Omega') \int dU \langle T_{kk} \rangle (\Omega'),
\end{equation}
\begin{equation}
\int \langle T_{U U}\rangle d U= - e^{-\kappa_+ d/2} \frac{c \kappa_+}{16 r_+^2},
\end{equation}
and
\begin{equation}
\label{eq:GFFourier}
H = \sum_{j} Y_{m=0, j}(\Omega) H_{mj}, \ \ \ H_{mj} =\sqrt{\frac{2 j+1}{ 4\pi}} \frac{2r_+^2}{2\kappa_+ r_+  + j(j+1)}
\end{equation}
for $Y_{m=0, j}(\Omega)  = \sqrt{\frac{2j+1}{4\pi}}P_j(\cos \theta)$ are standard scalar spherical harmonics on $S^2$ with vanishing azimuthal quantum number.

Putting this all together, we find
\begin{equation}
\Delta V = - \frac{ G_N c \pi \kappa_+}{2} \frac{1}{r_+ r_-} \left( \frac{r_+-r_-}{r_-} \right)^{-1-\left(r_-/r_+ \right)^2} e^{\kappa_+ \left(2r_+ - d/2 \right)} \int d \Omega' H(\Omega, \Omega').
\end{equation}
For concreteness, we can choose a geodesic at $\theta=\pi/2$, and keep just the lowest term in \ref{eq:GFFourier}, which dominates at small $\kappa_+$:
\begin{equation}
\Delta V = - \frac{ G_N c \pi }{2} \frac{1}{r_-} \left( \frac{r_+-r_-}{r_-} \right)^{-1-\left(r_-/r_+ \right)^2} e^{\kappa_+ \left(2r_+ - d/2 \right)} .
\end{equation}
%
where $c$ is the central charge associated with quantum fluctuations of the compact cosmic string, where we've chosen a geodesic through $\theta = \pi/2$, and where $U$ and $V$ are defined such that the metric on the bifurcation surface is
\begin{equation}
ds^2 = \frac{r_-}{r_+}\left(\frac{r_+ - r_-}{r_-}\right)^{1+(r_-/r_+)^2} e^{-2 \kappa_+ r_+} dU dV
\end{equation}
Then, if the time \eqref{shapiro} became longer than this available crossing time, a signal would be delayed by the presence of the third mouth for too long to make it across the wormhole. Considering the case that $b\sim r_e/2$, this leads to a bound
\begin{equation}
    \Delta v = \delta t \left(\frac{r_-}{r_+}\left(\frac{r_+ - r_-}{r_-}\right)^{1+(r_-/r_+)^2} e^{-2 \kappa_+ r_+}\right)^\frac{3}{2} \lesssim \Delta V
\end{equation}
which then gives
\begin{equation}
m \lesssim \frac{c \pi}{4 \log\left(\frac{4 d^2}{r_+^2}\right)} \frac{r_+^{1/2}}{r_-^{3/2}} \left( \frac{r_+-r_-}{r_-} \right)^{-3/2-3/2\left(r_-/r_+ \right)^2} e^{\kappa_+ \left(3r_+ - d/2 \right)}
\end{equation}
or, $m G_N \lesssim \frac{\ell_p^2}{r_-}$, which is suppressed relative to the construction in the Appendix \ref{Malda} by an additional factor of $\ell_p/r_-$, making it more difficult to create a semiclassical black hole in the throat.

\newpage

\bibliography{mbwh}
\bibliographystyle{utcaps}

\end{document}